\documentclass[12pt]{article}

\usepackage[caption=false]{subfig}% Support for small, `sub' figures and tables

\usepackage[natbibapa,nodoi]{apacite}% Citation support using apacite.sty. Commands using natbib.sty MUST be deactivated first!
\usepackage{amsmath}
\usepackage{pdflscape}
\usepackage{mathtools}
\usepackage{amsfonts}
\usepackage{array}
\usepackage{eucal}
\usepackage[ruled,lined]{algorithm2e}
\usepackage{dsfont}
\usepackage{longtable} 
\usepackage{float}
\usepackage{xcolor}
\usepackage{verbatim} % provides the comment function
\usepackage{lscape}  % allows for pages sidewise
\usepackage{enumitem}
\setlist[itemize]{align=parleft,left=0pt..1em}
\usepackage[margin=1.15in]{geometry}

\newcommand\blfootnote[1]{%
  \begingroup
  \renewcommand\thefootnote{}\footnote{#1}%
  \addtocounter{footnote}{-1}%
  \endgroup
}

\DeclareMathOperator*{\argmin}{arg\,min}

\begin{document}

\title{\textbf{Pooled Grocery Delivery with Tight Deadlines from Multiple Depots}}
\date{}
\author{
Maximilian Kronmueller$^{a,\star}$\blfootnote{CONTACT: M. Kronmueller Email: m.kronmuller(at)tudelft.nl},
Andres Fielbaum$^a$ and
Javier Alonso-Mora$^a$
}

\maketitle
\vspace{-0.65cm}
$^a$Delft University of Technology

\vspace{0.5cm}
\begin{abstract}
\noindent
\textbf{We study routing for on-demand last-mile logistics with two crucial novel features: i) Multiple depots, optimizing where to pick-up every order, ii) Allowing vehicles to perform depot returns prior to being empty, thus adapting their routes to include new orders online. Both features result in shorter distances and more agile planning.
We propose a scalable dynamic method to deliver orders as fast as possible. Following a rolling horizon approach, each time step the following is executed. First, define potential pick-up locations and identify which groups of orders can be transported together, with which vehicle and following which route. Then, decide which of these potential groups of orders will be executed and by which vehicle by solving an integer linear program. We simulate one day of service in Amsterdam that considers 10,000 requests, compare results to several strategies and test different scenarios. Results underpin the advantages of the proposed method.}
\end{abstract}

\vspace{0.5cm}

\section{Introduction} 
The possibility to order and have one's parcel delivered within the next minutes or hours, minimally within the same day, is appreciated by many customers. Thus, many young companies that offer \emph{flash deliveries} started to rise during the last years. Examples such as Gorillas, Flink, Getir, or GoPuff, are engaged in delivering last-mile grocery in minutes. In the Netherlands alone, consumers spent around 40 million euros per month on flash deliveries at the end of 2021, a trend that is continuously rising (\cite{Kantar_flash}). Even some supermarket brands are starting their first tries in these directions. For instance, a recent collaboration in the Netherlands between a supermarket brand (Albert Heijn) and food delivery brands (Thuisbezorgd and Deliveroo), is exploring delivering groceries as fast as possible (\cite{albert_thuisbezorgd}). Similarly, in several countries in South and North America, a delivery company (Cornershop) has recently merged with Uber for a similar purpose (\cite{cornershop_uber}).\\
Therefore, planning and routing algorithms are necessary to compute efficient vehicles' plans for such operations. Moreover, the progress in the area of autonomous delivery vehicles increases the importance of these algorithms even further, as it might become feasible to operate large fleets with moderate costs, and without the inherent risks that the human riders currently face (\cite{CHRISTIE2019115,Zheng_crash_risk,amiri2022adoption}).\\
This leads to planning for same-day delivery (SDD) operations or on-demand last-mile deliveries, serving customers at their homes. Most last-mile deliveries are operated using a single depot and with vehicles' routes planned and fixed when leaving the depot. This paper relaxes these two assumptions, proposing methods to choose the most convenient depot and to update the vehicles' itinerary while en route.

SDD can be described as follows: Orders are placed continuously throughout the day and need to be delivered before the end of the day or during a short time window. Parcels need to be picked up and delivered to customers' locations leveraging a fleet of vehicles. For each vehicle, a route needs to be found such that a given objective function is optimized, for example, maximizing the number of delivered packages or minimizing the waiting time of customers.

This paper studies a SDD problem aiming to deliver orders quickly. This is formalized using additional constraints, such as a short maximum delay. Further, we increase options to pick up orders from one depot to multiple. Moreover, most SDD operations assume that their vehicles deliver all loaded orders, afterwards return to the depot, and then become available to load and service new customers. We relax this assumption and allow for depot returns prior to being empty (pre-empty depot returns). This allows incorporating new occurring orders within existing plans of vehicles flexibly. We show that those additions can lead to more efficient routes and cheaper operations.

\begin{figure}[H]
	\centering
	\includegraphics[width=0.49\textwidth]{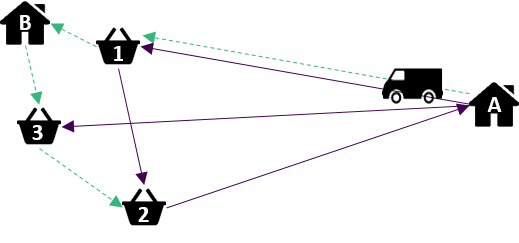}
	\caption{An exemplary tour of one vehicle serving two known orders (1 \& 2) and one newly requested order (3) that gets placed after the vehicle has already left depot A. The solid purple arrows show the route of the vehicle when there is only one depot and no pre-empty depot returns. The dashed green arrows show the route when using multiple depots (A \& B) and allowing for pre-empty depot returns.}
	\label{fig:intro}
\end{figure} 

Figure \ref{fig:intro} shows an example showcasing the difference between using multiple depots and allowing for pre-empty depot returns compared to a single depot and fixed routes after leaving the depot. Orders 1 and 2 are known and loaded into the vehicle. While the vehicle is on its tour a new order (order 3) occurs. If using depot A only and not allowing for pre-empty depot returns, the vehicle serves the two loaded orders, following the first part of the solid purple tour. Subsequently, it needs to return to the depot (depot A) and to drive to the new customer individually (the second part of the solid purple tour). If a second depot was available (depot B) and the possibility of depot returns prior to being empty was allowed, the original tour can be updated on the fly. The vehicle can load the new order at depot B after serving order 1, and can then service order 3 before serving customer 2 (dashed green tour). By doing so, the long way back to the depot can be saved and more efficient routes are possible. Further, customer 3 is served more quickly at the price of delaying order 2 slightly. As such, both operators and users can benefit.

We approach the above-described problem in a rolling horizon fashion, i.e., we divide the full-day problem into multiple subsequent sub-problems. The problem evolves with time as new orders arrive throughout the day. Further, the operation is planned (solving sub-problems) and executed (following the obtained routes) simultaneously, constantly changing the load and position of vehicles. To solve a single sub-problem at a specific time $t$, we first identify potential pick-up locations for each order. Second, potential feasible \textit{trips} are calculated, i.e., sequences to pick up goods and deliver orders. To assign these trips to vehicles an integer-linear program is solved. As a result, each vehicle has a constantly updated plan to follow, i.e., which orders to pick up and where, as well as in which sequence to deliver them.\\

\noindent The main contributions of this paper are threefold:
\begin{itemize}
    \item First, we consider multiple depots at which orders can be picked up. The method decides endogenously which depot to use for each order given the current state. Up to our knowledge, this is the first work that considers multiple depots for a dynamic vehicle routing problem without decomposing it into sub-problems, each having a single depot.
    \item Second, we allow vehicles to visit a depot to load additional parcels before distributing their loads, if this increases overall efficiency. 
    \item  Finally, our method can scale up to scenarios with thousands of orders and find good quality solutions online.
\end{itemize}
We evaluate the method by comparing it to a greedy algorithm, and to two scenarios that assess the relevance of the first two contributions described above: i) assuming that each order is picked up at its closest depot, and ii) prohibiting pre-empty depot returns. A comprehensive sensitivity study analyzes the effects of single parameters.\\

\section{Related Work} \label{sec:sota}
The same-day delivery problem can be categorized as a dynamic and possibly stochastic pick-up and delivery problem with incomplete information. Hence, it falls into the family of dynamic vehicle routing problems (DVRP), for which a comprehensive overview can be found in \cite{pillac_review_2013} and in \cite{psaraftis_dynamic_2016}.
Furthermore, \cite{berbeglia_dynamic_2010} provide an overview focusing specifically on dynamic pick-up and delivery problems. Additional related problems are the meal delivery routing problem (\cite{reyes_meal_2018, yildiz_provably_2019, ulmer_restaurant_2020}), vehicle routing to transport people (dial-a-ride problem) (\cite{alonso-mora_-demand_2017, cordeau_dial--ride_2007}), and the multi-robot task assignment problem (\cite{khamis_multi-robot_2015}). Further works focus on the integration of robots into routes of vans, for instance \cite{LIU2021102466}.
Note that we exclude detailed descriptions of works dealing with static vehicle routing problems (VRP) as the focus of the corresponding approaches differ. Instead, we refer interested readers to suitable review papers, such as \cite{VRP_book_Toth}.\\

The SDD literature splits into two: First, works focusing on vehicle dispatching or order acceptance followed by a separate routing step e.g., \cite{ghiani_anticipatory_2009, azi_dynamic_2012, klapp_one-dimensional_2016, klapp_dynamic_2018, klapp_order_2019, ulmer_same-day_2019} and second, work focusing on SDD routing directly. 
We look exclusively at the second part, including works like \cite{voccia_same-day_2017}, \cite{ulmer_preemptive_2018}, \cite{Kronmueller_MRS} and \cite{Cote2021Dynamic}. This line of research is more closely related to our approach, as routing decisions are at the core of the proposed method.\\

Let us look into the papers by \cite{voccia_same-day_2017}, \cite{ulmer_preemptive_2018} and \cite{Cote2021Dynamic} in more detail. \cite{voccia_same-day_2017} use a multi-scenario sampling approach, first introduced by \cite{MSA-Bent}. They are leveraging waiting strategies and test on scenarios with up to 800 orders and up to 13 vehicles. In contrast, our approach can handle larger problem sizes and allows to pick up orders at multiple depots, but works myopically.
Similar to our work, \cite{ulmer_preemptive_2018} allows for preemptive depot returns, i.e., depot returns before finishing the currently planned tour based on expectations of future events. The authors proposed a method that builds on approximate dynamic programming combined with an insertion routing heuristic. The method allows vehicles to return to depots before finishing their current routes. Our approach differs because pre-empty depot returns do not use anticipation of the unknown future but only use currently available information. The method by \cite{ulmer_preemptive_2018} can plan for a single vehicle, whereas our approach scales up to large fleet sizes.
\cite{Cote2021Dynamic} proposes different large neighborhood search based approaches for the SDD problem ranging from a re-optimization heuristic to a branch-and-regret heuristic. They rely on a multi-scenario approach to anticipate future events; and their approach is capable of performing preemptive depot returns as well. Algorithms were tested based on the same scenarios as \cite{voccia_same-day_2017}. Scenarios of up to ten vehicles were analyzed.\\

The multi-depot vehicle routing problem (MDVRP) is a static VRP featuring multiple depots and thereby introducing a choice, namely where to pick up the goods. Its dynamic counterpart is the dynamic multi-depot vehicle routing problem (DMDVRP), combining DVRP and MDVRP, which is the case we focus on in this paper. The static case has been thoroughly studied. \cite{montoya-torres_literature_2015} provides a good overview of the studies on this topic. For more details, we refer to their literature study and the associated papers.
Regarding DMDVRP, it has been tackled by decomposing the problem into multiple single-depot DVRPs, where each order is assigned to one fixed depot and each sub-problem is solved separately  (\cite{yu_improved_2013,xu_hybrid_2018}). In contrast, we include the decision of which depot should be used within the routing decision itself and thus this paper is the first, up to our knowledge, to consider multiple depots simultaneously for a DVRP.\\

The routing method proposed in this work is based on methods for ride-sharing. An overview of these methods can be found in \cite{agatz_optimization_2012, mourad_survey_2019, narayanan_shared_2020}. More specifically, our approach builds upon a routing and assignment method for a ridesharing system that transports people in metropolitan areas (\cite{alonso-mora_-demand_2017}). This method is called Vehicle-Group Assignment Method (VGA). VGA splits the procedure into two steps: First, it generates potential groups of orders that each vehicle can serve, and second, an optimal assignment of these potential groups to individual vehicles is computed. With realistic enough computation time, the method can solve large-scale real-world instances, up to thousands of vehicles, in an any-time optimal manner.
Our problem mainly differs in two aspects. The pick-up locations of orders are not pre-defined through themselves (nor optimized in a close area around them, as in \cite{fielbaum_-demand_2021}, who consider the option that passengers walk short distances), as any depot could be chosen. Hence, each order includes the additional decision where to pick it up. Second, the urgency of picking up an order fast is lower for delivering goods than for transporting people, as humans dislike waiting times. Therefore \cite{alonso-mora_-demand_2017} explicitly adds a constraint enforcing a maximum waiting time (i.e., time until pick-up) per customer. This can be omitted here. Additionally in practice, each vehicle can load more orders than people simultaneously without introducing any discomfort. Consequently, the number of feasible solutions in our scheme can increase due to having fewer constraints.\\
This work is an extension to the conference paper \cite{Kronmueller_MRS} and presents additional explanation, clarification and experiments.\\

\section{Problem Formulation} \label{sec:problem}
\subsection{Definitions}\label{sec:prob:def}
This section formally introduces and defines the various elements involved in the problem. The notation we use throughout the paper is summarized in Table \ref{tb:notation} in the Appendix.\\
\textbf{Environment:} Let $G = (\mathcal{N}, \mathcal{A})$  be a weighted directed graph where $\mathcal{N}$ defines a set of nodes and $\mathcal{A}$ defines a set of weighted arcs. The arcs' weights represent the travelling times between two connected nodes. We denote the shortest travel time between any two locations $x_1, x_2 \in \mathcal{N}$ by $\tau_{x_1,x_2}$. A depot $d \in \mathcal{N}$ is a specific node where goods can be picked up.
There are $\mathcal{H}$ depots in total, which are summarized in the set of depots $\mathcal{D} \subset \mathcal{N}$. We assume that every depot has all goods in stock. (We take this assumption for the sake of simplicity. However, it is straightforward how to extend our method if this wasn't the case.)\\
\textbf{Vehicle Fleet:} Vehicles can drive along the graph's arcs to load and deliver goods to customers. The amount of orders they can load is restricted through a maximum capacity C. The fleet $\mathcal{V}$ consists of $M$ identical vehicles. 
At each time $t$, a single vehicle $v \in \mathcal{V}$ is fully described by its current location $l_{v,t}$, and the orders it has loaded (picked-up and not yet dropped-off), summarized in the set $\mathcal{LO}_{v,t}$. The state of the whole fleet at time $t$ is denoted by $\mathcal{V}_t$.\\
\textbf{Demand:} The demand set is denoted by $\mathcal{O}$, where each order $o =(t_o, g_o) \in \mathcal{O}$ is revealed at time $t_o$ and has to be delivered to its destination/goal location $g_o \in \mathcal{N}$. We assume $t_o \in [0,T_{end}-\delta_T]$, where $T_{end}$ represents the end of the day and $\delta_T$ is a constant time span in which no more orders are placed. A total of $N$ orders are placed. For simplicity, we assume that all orders are of the same size, set to one. This assumption can easily be extended to variable order sizes. Note that an order itself does not define a pick-up location $p_o \in \mathcal{N}$.
At time $t$, the demand set $\mathcal{O}$ can be split into subsets depending on the status of each order $o \in \mathcal{O}$: 
\begin{itemize}
    \item The set $\mathcal{LO}_t$ consists of all orders $o \in \mathcal{O}$ that are currently loaded to any vehicle $v \in \mathcal{V}$, i.e., $\mathcal{LO}_t = \cup_{v \in \mathcal{V}} \mathcal{LO}_{v,t}$.
    \item The set $\mathcal{DO}_t$ consists of all orders $o \in \mathcal{O}$ that were delivered to their destinations $g_o$ before $t$.
    \item The set $\mathcal{IO}_t$ consists of all ignored orders that can not be delivered within given constraints anymore.
    \item The set $\mathcal{PO}_t$ consist of all orders $o \in \mathcal{O}$ that are already known (i.e. $t_o \leq t$) but have not been picked-up, delivered or ignored yet.
    \item For completeness, $\mathcal{UO}_t$ is the set of all unknown orders, consisting of all orders $o \in \mathcal{O}$ such that $ t_o > t$.
\end{itemize}

The subsets are defined such that each order only belongs to one subset at time $t$, thus they are disjoint, and fulfill $\mathcal{O} = \mathcal{UO}_t \cup \mathcal{PO}_t \cup \mathcal{LO}_t \cup \mathcal{DO}_t \cup \mathcal{IO}_t$. At the beginning of the day ($t=0$), all orders are unknown, i.e. $\mathcal{UO}_0 = \mathcal{O}$. At the end of the day ($t=T_{end}$), all orders are either delivered or ignored, i.e., $\mathcal{DO}_{T_{end}} \cup \mathcal{IO}_{T_{end}} = \mathcal{O}$ and $\mathcal{UO}_{T_{end}} = \mathcal{PO}_{T_{end}} = \mathcal{LO}_{T_{end}} = \emptyset$.

\begin{figure}[H]
	\centering
	\includegraphics[width=0.7\columnwidth]{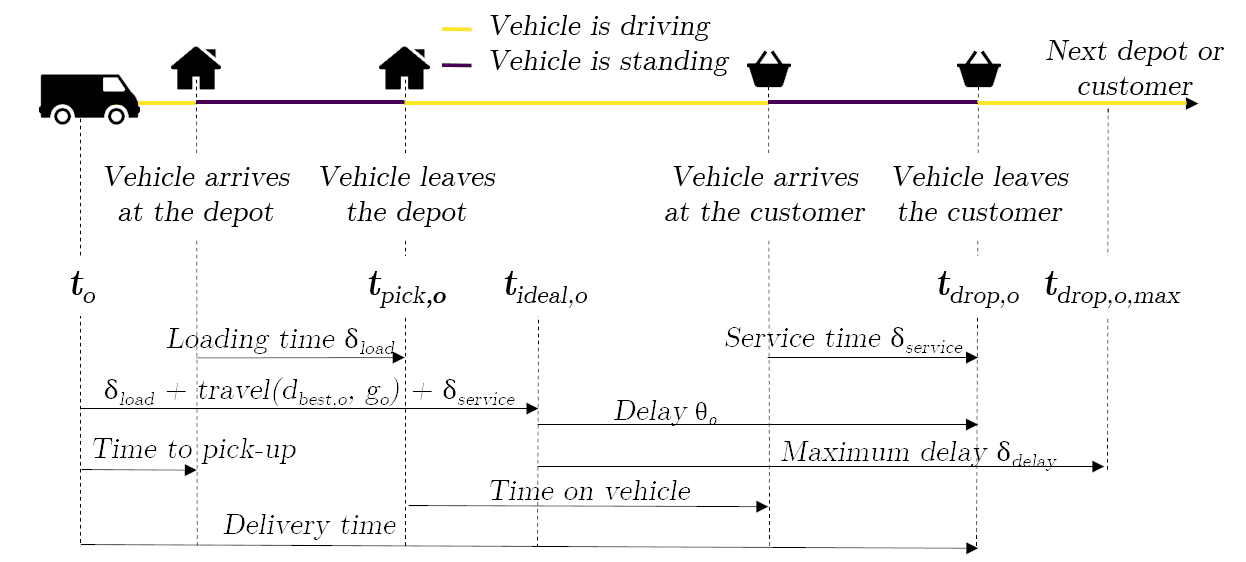}
	\caption{Visualization of the different times and time spans for one order.}
	\label{fig:times}
\end{figure}

\textbf{Times:} We assume that vehicles need some constant time to load or deliver a single order, denoted by $\delta_{load}$ and $\delta_{service}$, during which they are standing still.
The times at which an order $o$ is picked up and dropped-off are denoted by $t_{pick,o}$ and $t_{drop,o}$ respectively. The earliest time an order can be delivered is described by $t_{ideal,o}$. To do so, an idle vehicle needs to be located at the closest depot to the order's destination $d_{\text{best},o}$, and start serving the customer immediately without any detours, resulting in $t_{ideal,o} = t_o + \delta_{load} + \tau_{d_{\text{best},o},g_o} + \delta_{service}$. This is a lower bound for the delivery time, i.e., $t_{drop,o} \geq t_{ideal,o}$. We define the difference between the ideal and actual delivery time as the delay $\theta_o = t_{drop,o} - t_{ideal,o} \geq 0$. Further, we assign each order a maximal drop off time $t_{drop,o,max} = t_{ideal,o} + \delta_{delay}$, where $\delta_{delay}$ is the maximally allowed delay per order, and is predefined by the operator to ensure a desired service level.
A summary of all involved points in time for one order is illustrated in Figure \ref{fig:times}.

\subsection{Problem Description}
\subsubsection{Full-Day Problem} \label{full_day_problem}
The problem describing the whole day can be posed as follows:\\

\textit{Consider a directed and weighted graph $G=(\mathcal{N},\mathcal{A})$, a set of depots $\mathcal{D}\subset \mathcal{N}$ where orders can be picked up, a fleet of autonomous vehicles $v \in \mathcal{V}$, each of them in its initial location $l_{v,0}$, and a demand $\mathcal{O}$. The operation starts at $t=0$ and ends at $t=T_{end}$, and each order $o \in \mathcal{O}$ is revealed at $t=t_o \in [0,T_{end}-\delta_T]$. The problem consists in finding an assignment $\Omega$ of orders $o \in \mathcal{O}$ to vehicles $v \in \mathcal{V}$, including the choice of a depot $d \in \mathcal{D}$ to pick-up each order and routes for each vehicle to follow, in order to minimize a given cost function $\mathcal{J}$ subject to a set of given constraints.}\\

The plan that a vehicle $v$ takes during the whole day ($[0,T_{end}]$) is denoted by $P_v$, and is given by the orders it serves, where and when to pick them up, and in which sequence they are served. Therefore, $\Omega = \cup_{v \in \mathcal{V}} P_v$. Further, each vehicle's plan needs to obey the following constraints.\\
\begin{itemize}
	\item Each vehicle has a maximum capacity of C $$\mathcal{LO}_{v,t} \leq C \quad \forall v \in \mathcal{V}, \; t \in [0,T_{end}]$$
	\item Each order is allowed to have a maximum delay of $\delta_{delay}$ otherwise the order is ignored
	$$\theta_o \leq \delta_{delay} \quad \forall o \in \mathcal{O} \setminus \mathcal{IO} $$ 
\end{itemize}

Orders that can not be delivered within the given constraints are ignored, i.e., they are not served.
In our simulations, we assume that at the beginning of the day ($t=0$) all vehicles $v \in \mathcal{V}$ are distributed over all depots $d \in \mathcal{D}$ and are empty $\mathcal{LO}_{v,0}=\emptyset \quad \forall v \in \mathcal{V}$. 
We use a \emph{cost function} $\mathcal{J}$ that takes the operator's and customers' costs into account. The customer's cost of order $o \in \mathcal{O}$ is defined as its delay $\theta_o$, so that it measures the quality of service. Thus the faster an order is delivered, the better. The operator's costs are defined as the sum of the travelling time of each vehicle $\tau_{v} $ to serve all orders assigned to it, expressed mathematically as $ \sum_{v \in \mathcal{V}} \tau_{v}$.
The two costs are combined convexly, via a weight $\beta$, called cost weight. A penalty $\alpha$ is charged if an order is ignored. The penalty $\alpha$ can be interpreted as the cost the operator has to cover if a third party is hired to deliver the respective order.
In this work we set $\alpha$ to be considerably larger than the sum of the other two cost terms, meaning that the system first aims at maximizing the number of served orders, and then to minimize the combination of operators' cost and customers' cost. 
The overall objective function at $t=T_{end}$ is represented by Equation \ref{eq:cost_full}.
\begin{equation} \label{eq:cost_full}
\mathcal{J}_{T_{end}} =  \left[ (1-\beta) \cdot \sum_{o \in \mathcal{DO}_{T_{end}}} \theta_o + \beta \cdot \sum_{v \in \mathcal{V}} \tau_{v} +  \sum_{o \in \mathcal{IO}_{T_{end}}} \alpha\right]
\end{equation}
It is worth commenting that the method we propose in Section \ref{sec:method} does not depend on this specific cost function; in other words, a different cost function could be used and our method still applies.\\
In all, our problem combines several NP-hard problems, including the capacitated vehicle routing problem (\cite{Ralphs-CVRP}), and the multi-depot vehicle routing problem (\cite{MONTOYATORRES-MDVRP}). Moreover, it requires dynamic optimization, and admits large fleet sizes.

\subsubsection{Problem Dynamics} \label{sec:problem_dynamic}
\begin{figure}[H]
	\centering
	\includegraphics[width=0.95\columnwidth]{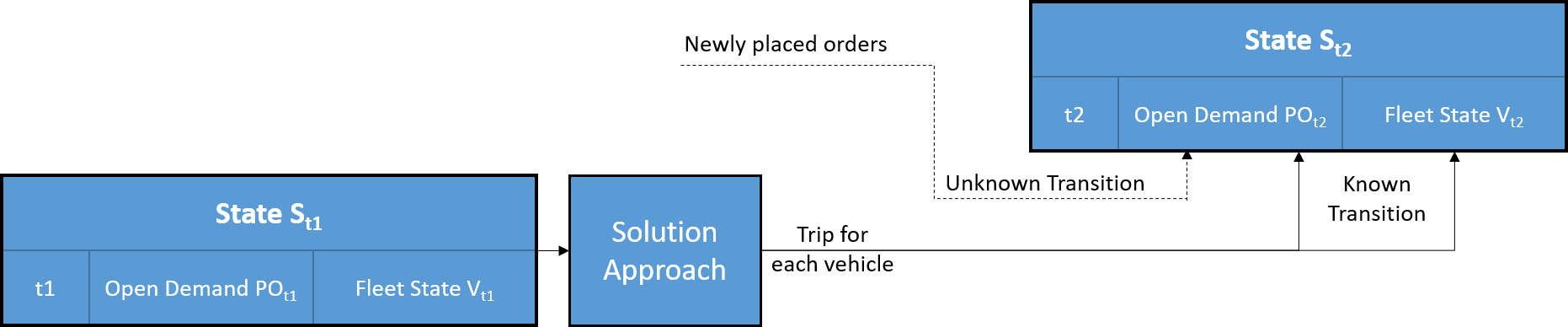}
	\caption{Visualization of the dynamics of the transition between two consecutive states. The transition of the vehicle fleet is known. On the other hand, the transition of the open demand is partly unknown due to customers placing new orders.}
	\label{fig:state_transition}
\end{figure}

The full-day problem description (Section \ref{full_day_problem}) does not cover the dynamics of the problem explicitly. The problem can be classified as dynamic because requests appear as they are served; thus, all decisions must be updated constantly. To do so, we first need to be able to describe the state of the problem at any time $t$. The problem state $\mathcal{S}_t$ at time $t$ is fully characterized by the time itself, the vehicles fleet state $\mathcal{V}_t$, and the set of known but not yet loaded orders $\mathcal{PO}_{t}$ at $t$. This results in the state definition as $$\mathcal{S}_t= (t,\mathcal{V}_t,\mathcal{PO}_t) \text{.}$$
The transition from a current state $\mathcal{S}_{t1}$ to a future state $\mathcal{S}_{t2}$ can be split into two. On one side, the transition of the vehicle fleet's status $\mathcal{V}_t$. This transition is known and only determined by the requests each vehicle is assigned to and the sequence to serve them. On the other side, the transition of the set of known but not yet loaded orders $\mathcal{PO}_{t}$ is determined by the requests assigned to all the vehicles, as some orders may get loaded during this time, and unknown exogenous information as customers placed new orders. We assume to have no knowledge about these future orders and also do not include any predictions about them. Figure \ref{fig:state_transition} depicts an schematic visualisation.

\section{Method} \label{sec:method}
This section gives a short overview of the proposed method and subsequently explains each component in detail. 

\subsection{Method Overview}
We tackle the introduced problem in a rolling horizon fashion, meaning that a decision for a given state is taken, then time is propagated forward (vehicles follow their plans and new orders get accumulated) to the point at which a new decision state is meet. 
This process is repeated until the end of the day. We propagate time in fixed time steps of duration $\Delta t$. This means that our approach is ``batch-based'', in which a number of requests are accumulated before deciding how to assign, as opposed "event-based" approaches, where each request is assigned as soon as it appears. The extra information allows to making more efficient decisions, as has already been acknowledged by the industry (\cite{UberWebsite}).)
A routing decision, enumerated by $k$, is done at time $t_k = k \cdot \Delta t$. This results in $\mathcal{K}=T_{end}/\Delta t$ decisions from start to end of the operation. A single decision $k$ at $t_k$ is the assignment of currently open orders to vehicles and a corresponding update on vehicles' itineraries, considering the corresponding state $S_{t_k}$, and optimizing the assignment according to the objective function. Solving this assignment $\mathcal{K}$ times consecutively gives a solution for the overall problem formulated in Section \ref{full_day_problem}.\\
Our approach is myopic, i.e., it does not take future states into account. We regard this assumption as reasonable (not optimal), as $\Delta t$, the time between two consecutive decisions, is rather short (100 seconds in our experiments). Thus, new information is included to the problem fast and previous solutions are updated frequently. Further, myopic approaches are usual in the scientific literature, although anticipatory techniques can be used to improve the solutions, which is regarded as a relevant direction for future work.
For a discussion on this topic, see \cite{Ulmer_off_on,MSA-Bent,FielbaumAnticipation,hyland2020integrating}. 

\begin{figure}
	\centering
	\includegraphics[width=1\columnwidth]{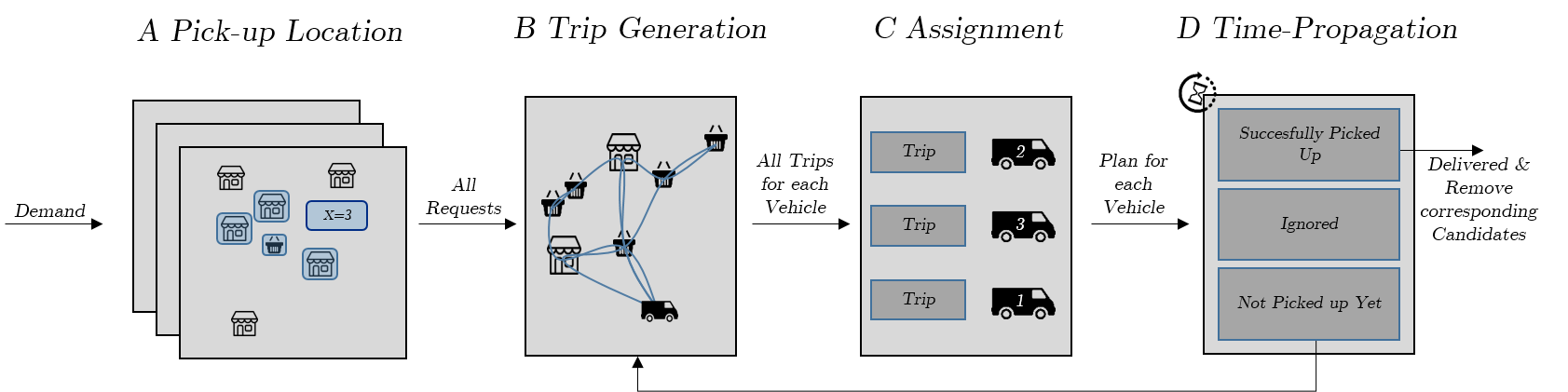}
	\caption{Schematic overview of our solution approach. Step A assigns several potential pick-up locations to each order. During step B, individual candidates (combinations of orders and specific pick-up locations) are combined to feasible trips. In Step C, trips to be executed and corresponding vehicles are selected. Within step D, we propagate time and vehicles follow their assigned plans.}
	\label{fig:Solution_Approach_Overview}
\end{figure}

To solve one specific state of the problem, we propose a method divided into four steps: First, potential pick-up locations for each order are found. Second, orders with associated pick-up locations are grouped into potential trips, taking the current location of each vehicle into account. A trip is an ordered sequence of locations to pick up and deliver orders executed by a vehicle. With enough computational time, we calculate all possible trips for each vehicle. Third, we decide which of these potential trips are being executed. Last, vehicles execute parts of their plans as time is propagated forward. 
An overview of the approach is depicted in Figure \ref{fig:Solution_Approach_Overview}.

\subsection{Finding Pick-up Locations} \label{sec:method_pickup}
Each individual order $o \in \mathcal{O}$ needs to be assigned to a specific pick-up location $p_o \in \mathcal{D}$. A depot $d \in \mathcal{D}$ is a feasible option for an order $o$, if a vehicle can pick up the goods at $d$ and delivery them in time. Each order might have more than one feasible depot. To select one of these options, we first define the term \textit{candidate} $c$ of an order $o \in \mathcal{O}$ as follows.\\

\textbf{\textit{Definition:}} A candidate $c$ is a tuple containing an order $o_c \in \mathcal{O}$ and an associated pick-up location $p_c \in \mathcal{D}$. Thus a candidate is described as $c=(o_c,p_c)$.\\

A candidate $c$ is unique, but one order $o \in \mathcal{O}$ can have multiple candidates associated with it, each having a different pick-up location $p_c \in \mathcal{D}$.
$\mathcal{I}^{\mathcal{C}}_o$ denotes the set of candidates that belong to order $o \in \mathcal{O}$. The set of all candidates is denoted by $\mathcal{C}$. $\mathcal{C}_t$ is the set of candidates at time $t$ corresponding to all placed orders $o \in \mathcal{PO}_t$.\\
We introduce a heuristic to select a subset of pick-up locations. We consider the $x$ depots closest to the destination measured in travel time. The parameter $x$ can be tuned. This results in maximally $x$ candidates per placed order.

If $x = H$ all feasible depots are considered and if $x = 1$ only the closest depot is used for each order.
We do so to control the number of candidates per order and thus the number of potential trips for each vehicle, which is directly correlated to the required computational effort.\\
If $x=1$, i.e. a single depot is considered per order, this approach resembles a decomposition of the full problem into multiple single-depot problems (an approach explained in Section \ref{sec:sota})
; however, in decomposition approaches vehicles are fixed to one depot, which is more restrictive than our approach even if we use $x=1$.

\subsection{Trip Generation} \label{sec:method_trip}
\begin{figure}[H]
	\centering
	\includegraphics[width=0.95\textwidth]{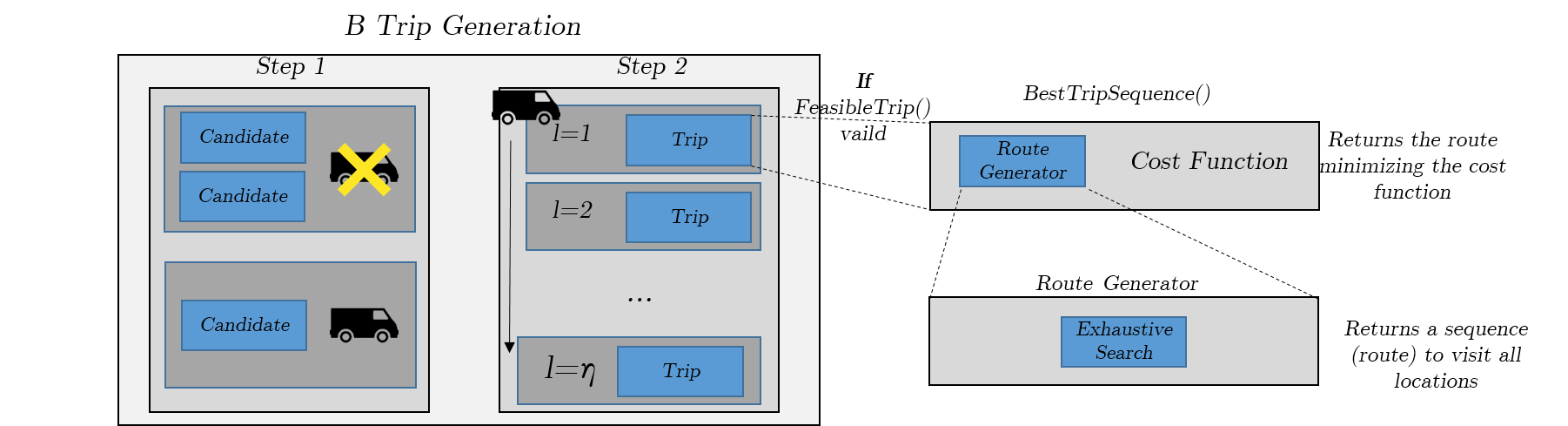}
	\caption{Schematic overview of the trip generation step. In step 1, we check the pairwise feasibility of combining two candidates and if one candidate can be combined with one vehicle into a trip. During step 2, trips are increased in size in steps of one until no more candidates can be added or a predefined maximum $\eta$ is reached. To do so, it needs to be checked if a trip is feasible in the first place and which is the best route to deliver it.}
	\label{fig:trip_gen_Overview}
\end{figure}

We calculate the set of feasible trips for each vehicle in the trip generation step. A trip $T$ is defined as an ordered sequence of locations to pick up and deliver candidates to be executed by a vehicle. To calculate potential feasible trips for a vehicle, we look at sets of candidates separately for each vehicle.
An overview of the trip generation process is depicted in Figure \ref{fig:trip_gen_Overview}.
This trip generation process is done iteratively, it starts by calculating small trips. We do so to leverage the idea that a trip can only be feasible, if all its sub-parts are feasible as well. A trip's size $l$, measured as the number of considered candidates, is thereby step-wise increased starting at a size of one until a maximum size $\eta$ is reached. The operator sets $\eta$. Additionally, huge trips are prevented as each order has a latest drop-off time $t_{drop,o,max}$. The result of this step is a set of potential trips for each vehicle. The trip definition contains a set of candidates $c$ to be delivered in the trip's route, and thus a set of orders that will be served.
The algorithm to calculate the set of all feasible trips $\mathcal{T}_t$ at time $t$ (except for candidates that were discarded due to the parameter $x$), is shown in Algorithm \ref{algo:trips}.\\
\begin{algorithm}
	\caption{Trip Generation for decision $k$ at $t_k$}
	\label{algo:trips}
	\SetKwInOut{Input}{input}
	\SetKwInOut{Output}{output}
	\Input{$S_{t_k}, \mathcal{C}_{t_k}, \eta$}
	\Output{All feasible trips $\mathcal{T}_{t_k}$}
	\SetKwBlock{Beginn}{beginn}{ende}
	\Begin{
		$\mathcal{T}_{t_k} = \emptyset$ \;
		\ForEach{$ v \in \mathcal{V}$}{
			$\mathcal{T}_\ell = \emptyset \quad \forall \ell \in \{1,..., \eta \} \;	\text{\scriptsize  (Set of all trips of size $\ell$)}$\;
			[add trips of size 1] \\
			\ForEach{$ c \in \mathcal{C}_{t_k}$}{
			    \If{$CandidateVehicle(v,c)$ valid}{
				$\mathcal{T}_1 \leftarrow \mathcal{T}_1 \cup (c)$ \, \scriptsize (Add trip to set of trips)}}
			[add trips of size 2] \\
			\ForEach{$(c_i)$, $(c_j) \in \mathcal{T}_1$ }{\If{$TwoCandidates(c_i,c_j)$ valid and $FeasibleTrip(v,c_i,c_j)$ valid}{
				$\mathcal{T}_2 \leftarrow \mathcal{T}_2 \cup BestTripSequence(v,c_i,c_j)$}}
			[add trips of size $\ell$] \\
			\For{$\ell \in \{3,..., \eta \}$}{
				\ForEach{$T_i,T_j \in \mathcal{T}_{\ell-1}$ with $|T_i \cup T_j|=\ell$ \, \text{\scriptsize (The two combined trips contain $\ell$ candidates together)}}{
					\If{$\forall h \in \{1,...,\ell\}, \{c_1,...,c_\ell\} \setminus c_h \in \mathcal{T}_{\ell-1}$ \, \text{\scriptsize (Each subset of this trip is a feasible smaller trip)}}{\If{$FeasibleTrip(v,T_i \cup T_j)$ valid}{
							$\mathcal{T}_\ell \leftarrow \mathcal{T}_\ell \cup BestTripSequence(T_i \cup T_j)$\;}}}}
		}%\EndFor
		\Return $\mathcal{T}_{t_k} \leftarrow \cup_{\ell \in \{1,..., \eta \}}\mathcal{T}_\ell$
	}
\end{algorithm}
In Algorithm \ref{algo:trips} we use four functions: $CandidateVehicle()$, $TwoCandidates()$, $FeasibleTrip()$ and $BestTripSequence()$, each explained in detail in the following:
\begin{itemize}
    \item The binary logic function \textbf{\textit{CandidateVehicle}($v,c$)} is valid if vehicle $v$ can serve candidate $c$ in the current state of the problem without violating any constraint. Therefore, we check if the vehicle can drive from its current location to the candidate's pick-up location and can deliver the order to its destination without violating the constraints of the order and the constraints of the other parcels on board as well as the vehicle's maximal capacity constraint.
    \item The binary logic function \textbf{\textit{TwoCandidates}($c_i,c_j$)} checks whether the two candidates $c_i$ and $c_j$ are combinable, i.e., if they can both be served by a hypothetical vehicle located at the corresponding depot satisfying all the constraints. As multiple candidates per order exist, we add a constraint to the existing time and capacity constraints: For two candidates to be combinable into one trip, we require them to share their pick-up location. Note that if a vehicle is at a depot and will serve two orders, it is more time-efficient to pick up both now rather than at two different depots at different times. This also implies that two combinable candidates cannot belong to the same order. (Let us analyze the effect of adding the constraint that two candidates need to share their pickup location to be combinable in more detail. For $q$ orders with $q$ different pick-up and $q$ different drop-off locations $\frac{(2q)!}{2^q}$ possible sequences to visit them exist. For $q$ orders that share their pick-up location this decreases to $q!$, where it is easy to see that $\frac{(2q)!}{2^q} > q!$ for $q > 1$. To estimate the size of the effect of these two conditions, we divide their respective number of possibilities, resulting in a factor of $\frac{2^{-q}(2q)!}{q!}$. This factor rises quickly; for five orders ($q=5$), we already get a factor of 945 times the number of potential sequences. Thus, this effect quickly outweighs the complexity of adding multiple candidates per order.)
    \item The binary logic function \textbf{\textit{FeasibleTrip}($v,T$)} checks weather all orders of a trip $T$ can be feasibly served by the vehicle $v$, i.e., whether it is able to go from its current location to serve all the candidates of the trip without violating any of the constraints.
    \item If a trip $T$ is feasible, we determine the sequence in which to deliver all its candidates using the function \textbf{\textit{BestTripSequence}($T$)}. Sometimes there will be more than a single feasible route to visit all trip locations. The function $BestTripSequence(T)$, chooses the route that minimizes the given cost function for the trip, explained below (Equation \ref{eq:gamma}). We perform an exhaustive search over all possible sequences of the trip. For large group sizes, some routing heuristics (such as insertion heuristics) could be used.
\end{itemize}
The cost of visiting a sequence of locations associated to trip $T$ and vehicle $v$ is given by $\gamma_{T,v}$, which is derived from Equation \ref{eq:cost_full}, the overall cost function:
\begin{equation} \label{eq:gamma}
\gamma_{T,v} : = (1-\beta) \cdot \sum_{o \in T} \theta_o + \beta \cdot \tau_T
\end{equation}
where $\tau_T$ represents the total driven distance to complete trip $T$. 
For vehicles that already contain load, the sequence includes those loaded orders. The relative position within the trip's route in which the prior loaded orders and new ones are served is unconstrained. Herein the possibility of pre-empty depot returns occurs.
We only keep the route (sequence of visited locations) that minimizes the costs (Equation \ref{eq:gamma}) of the trip for a specific vehicle and a set of candidates. Taking the minimal cost route is included in the subsequent notation of a trip $T$. Calculations for one vehicle are stopped if a predefined time, $\rho_{max}$, has passed. In this case, the trips generated up to this point are considered.\\

\subsection{Assignment of Trips to Vehicles} \label{sec:method_ilp}
After calculating the set of potential feasible trips $\mathcal{T}_t$ (previous step), we need to decide which of them should be carried out to minimize the objective function (Equation \ref{eq:ILP_1}). We call this step the assignment. The assignment is formulated as an integer linear program (ILP). 
The ILP is presented in Algorithm \ref{alg:opt}.
\begin{algorithm}
	\caption{Assignment}
	\label{alg:opt}
	\SetKwInOut{Input}{input}
	\SetKwInOut{Output}{output}
	\Input{Greedy assignment of trips to vehicles $\Omega_{\text{init}}$}
	\Output{Assignment of trips to vehicles $\Omega_{\text{optim}}$}
	\SetKwBlock{Beginn}{beginn}{ende}
	\Begin{
		Initalize with $\Omega_{\text{init}}$ \;
		Solve\;
		\begin{equation} \label{eq:ILP_1}
		\Omega_{\text{optim}} = \argmin_{\chi} \sum_{T,r \in \epsilon_{\mathcal{T}\mathcal{V}}} (\gamma_{T,v}-\gamma_{loaded,v}) \epsilon_{\mathcal{T},v} + \sum_{o \in \{1, ...,|\mathcal{PO}_t|
			\}} \alpha\chi_o
		\end{equation}
		
		\begin{equation} \label{eq:ILP_2}
		\sum_{T \in \mathcal{I}^T_{v}} \epsilon_{\mathcal{T},v} \leq 1  \quad \quad \quad \forall v \in \mathcal{V}
		\end{equation}
		\begin{equation} \label{eq:ILP_3}
		\sum_{c \in \mathcal{I}^{\mathcal{C}}_{o}} \sum_{T \in \mathcal{I}^T_{c_o}} \sum_{v \in \mathcal{I}^{\mathcal{V}}_{T}} \epsilon_{\mathcal{T},v} + \chi_o = 1 \quad \forall o \in \mathcal{PO}_t
		\end{equation}	
		\begin{equation} \label{eq:ILP_4}
		\chi_o \in \left\lbrace 0,1 \right\rbrace 
		\end{equation}
		\begin{equation} \label{eq:ILP_5}
		\epsilon_{\mathcal{T},v} \in \left\lbrace 0,1 \right\rbrace 
		\end{equation}	
		\Return $\Omega_{\text{optim}}$
	}
\end{algorithm}
Thereby, $\epsilon_{\mathcal{T}\mathcal{V}}$ denotes the set of all feasible trip vehicle combinations, and $\epsilon_{\mathcal{T},v}$ is the corresponding binary variable, taking the value 1 if the combination is executed.
Further, we create the following sets: $\mathcal{I}^T_{v}$, the set of trips that can be serviced by a fixed vehicle $v \in \mathcal{V}$; $\mathcal{I}^T_c$, the set of trips that contain candidate $c$; $\mathcal{I}^{\mathcal{V}}_{T}$, the set of vehicles that can service trip $T$; $\mathcal{I}^{\mathcal{C}}_{o}$, the set of candidates that belong to order $o$. Further, $\chi_o$ is a binary variable, taking the value of one if the corresponding order is ignored, and $\mathcal{X}$ is a set of all variables $\mathcal{X} = \{\epsilon_{\mathcal{T},v} , \chi_o ; \, \forall \epsilon_{\mathcal{T}\mathcal{V}} \text{ and } \forall o \in \mathcal{O} \}$.\\
Equation \ref{eq:ILP_1} describes the objective function for a single state. Note that the considered costs are relative. From the costs of a vehicle's route $\gamma_{T,v}$ (see Equation \ref{eq:gamma}), the costs for the considered vehicle to serve its already loaded parcels are subtracted, $\gamma_{loaded,v}$. Thus, we only account for changes in the vehicle's plan. If a vehicle's plan is not changed by not assigning any new orders, the assignment poses no costs. Note that this does not affect the optimization according to the global objective function. Equation \ref{eq:ILP_2} ensures that each vehicle is at most assigned to one trip. Equation \ref{eq:ILP_3} ensures that each order is assigned to a single vehicle or is rejected and the penalty $\alpha$ is charged. Furthermore, it ensures that no more than one candidate belonging to the same order is chosen. Equations \ref{eq:ILP_4}-\ref{eq:ILP_5} ensure that the corresponding variables are binary. $\chi_o$ takes the value one if its associated order $o \in \mathcal{O}$ can not be served by any vehicle or is ignored. Equation \ref{eq:ILP_5} defines $\epsilon_{\mathcal{T},v}$ as binary. As a result, each vehicle is assigned to a new trip or does not receive any new orders. If a vehicle receives no new orders, it will follow its current plan of delivering the currently loaded parcels or be considered idle if it has none.\\
To fasten the time needed to solve the above-presented ILP, we initialize it by a greedy solution.
The greedy solution is constructed by selecting the largest trip, measured by the number of served candidates $l$, first. If multiple trips serve the same amount of candidates, the trip with the lowest cost is selected.
We remove all trips which include already assigned orders or vehicles. We iterate until there are either no more vehicles or no more orders to assign. The pseudo-code of this initialization algorithm can be found in Appendix \ref{alg:greedy_initalize}.

\subsection{Return of Idle Vehicles}
If a vehicle is considered idle after an assignment, we instruct it to move towards the closest depot from its current location. We do so to enable the vehicle to pick up orders quickly in the following steps. Nevertheless, it may still be assigned otherwise in a future time step before reaching that depot.

\subsection{Time-Propagation} \label{sec:time_propagation}
In this step, we propagate time and update all elements affected by it, until the next decision $k+1$ is triggered, $t_{k+1} = t_k + \Delta t$. Each vehicle follows its plan determined in the previous steps (the trip assigned to it). Each order can take one of the following five states:
\begin{itemize}
	\item An order can be \textbf{picked up} by a vehicle at a depot ($o \rightarrow \mathcal{LO}_{t_{k+1}}$). As soon as an order is picked up it is fixed to this vehicle. It can not be unloaded anymore, meaning that it can not be reassigned to any other vehicle. Additionally, as multiple candidates belonging to one order exist, but only one of them gets served, the other candidates belonging to this order are removed.
	\item An order can be \textbf{delivered} to its destination ($o \rightarrow \mathcal{DO}_{t_{k+1}}$).
	\item An order can be assigned to a trip, but the planned pick-up time is later than $t_{k+1}$, the time of the next decision. Thus we consider the order as \textbf{not picked up}, yet. All not picked up orders, more precisely the associated candidates, are reinserted into the trip generation step for the next decision, thus allowing for reassignment ($o \rightarrow \mathcal{PO}_{t_{k+1}}$).
	\item An order can be assigned to no vehicle. This order (associated candidates) is reinserted into the trip generation step for the next decision ($o \rightarrow \mathcal{PO}_{t_{k+1}}$), unless it is no longer feasible to serve it as explained in the next bullet point.
	\item An order can be \textbf{ignored} ($o \rightarrow \mathcal{IO}_{t_{k+1}}$), meaning that it is not feasible to deliver it without violating any constraint. All candidates belonging to this order are removed.
\end{itemize}
Note that an order $o \in \mathcal{O}$ is ignored in the case it can't be delivered before the latest drop-off time $t_{drop,o,max}=t_{ideal,o}+\delta_{delay}$. Hereby, $t_{drop,o,max}$ is mainly influenced by the value of $\delta_{delay}$. The smaller $\delta_{delay}$ is set, the harder it is to combine multiple candidates to be served by one vehicle. On the other hand, if $\delta_{delay}$ is set too large, the number of possible combinations becomes vast, which can hinder solving the problem in the first place due to an increased combinatorial size. A good balance has to be found by the operator. We distinguish between $\delta_{\text{delay,real}}$, defined by the service level and $\delta_{\text{delay,heuristic}}$, the maximum delay at which the method performs well. In case that $\delta_{\text{delay,heuristic}} < \delta_{\text{delay,real}}$, the former should be used. To adjust to $\delta_{\text{delay,real}}$ we allow a candidate to be reinserted into the problem after it has been ``ignored''. The candidate gets reinserted with a new release time of $t_k$, the current time. Each candidate can be ignored up to a limit of $\zeta$ times, which is defined as:
\begin{equation} \label{eq:reinsert}
    \zeta = (\delta_{\text{delay,real}} -(\delta_{\text{delay,real}} \mod \delta_{\text{delay,heuristic}})) / \delta_{\text{delay,heuristic}} \text{.}
\end{equation}
When a candidate gets ignored $\zeta$ times, it is removed from the problem. Note that for feasibility calculations, the new release time has to be used. Nevertheless, the original release time is used to calculate the users' costs of a candidate on a trip.\\

\subsection{Complexity and Optimality Analysis} \label{sec:complexity_ana}
\textbf{Complexity:} Our approach divides the full-day problem (Section \ref{sec:problem}) into multiple sub-problems at specific times $t_k$. Each sub-problem deals with it's associated state $S_{t_k}$.
The trip generation step (Section \ref{sec:method_trip}) is the most complex and thus the bottleneck of the proposed approach. The ILP (Section \ref{sec:method_ilp}) can become large but stays solvable in a reasonable time by state-of-the-art solvers. Thus we analyze the trip generation step in more detail.\\
Let us do a worst-case scenario analysis, where all the orders are associated with the same $x$ depots, the corresponding candidates are all combinable, and all sets of candidates can be served by any vehicle. Recall that the maximum trip size is $\eta$. This leads to a complexity of:
$$\mathbb{O}(|\mathcal{V}| \cdot |\mathcal{O}|^\eta \cdot x)$$
If the trips' size become large, limited by $\eta$, the complexity can increase rapidly. In practice, the trip size is further influenced by two other factors:
First, the density of orders, i.e. the relation of the spatial size of the graph and the size of the set of orders, which affects how orders can be combined. The lower the density of orders is, the harder it becomes to serve them together. As a result, the maximum trip size decreases. Second, a short maximum delivery time also decreases the maximum length of potential trips and also their number.\\

\textbf{Optimality:} The proposed approach is able to solve a sub-problem, regarding a single state, to optimality. To achieve this, all depots have to be considered ($x=\mathcal{H}$), enough computational time has to be given, and the maximum trip length has to be unconstrained.
Note that even if each sub-problem is solved exactly, this does not imply an optimal solution to the full-day problem, due to the myopic approach employed.

\section{Experiments and Results} \label{sec:results}
This section first analyzes one simulation run in detail, representing a day of on-demand grocery delivery in Amsterdam (Section \ref{sec:result_base}), where we are able to deal with thousands of requests. To study the performance and benefits of the proposed method, we evaluate the approach by comparing it with a greedy approach, a scenario that considers a single depot per order, and a scenario that does not allow for pre-empty depot returns, (Section \ref{sec:result_compare}). In Section \ref{sec:result_sensitivity} we present the results of a sensitivity analysis of the main parameters. Table \ref{tb:results_numbers} in the Appendix contains all results of all analyzed scenarios in terms of precise numbers.

\subsection{Base Scenario: One Scenario in Detail} \label{sec:result_base}
\begin{figure}
\centering
\subfloat[Graph representing the city centre of Amsterdam.]{%
\resizebox*{7cm}{!}{\includegraphics[width=1.\textwidth]{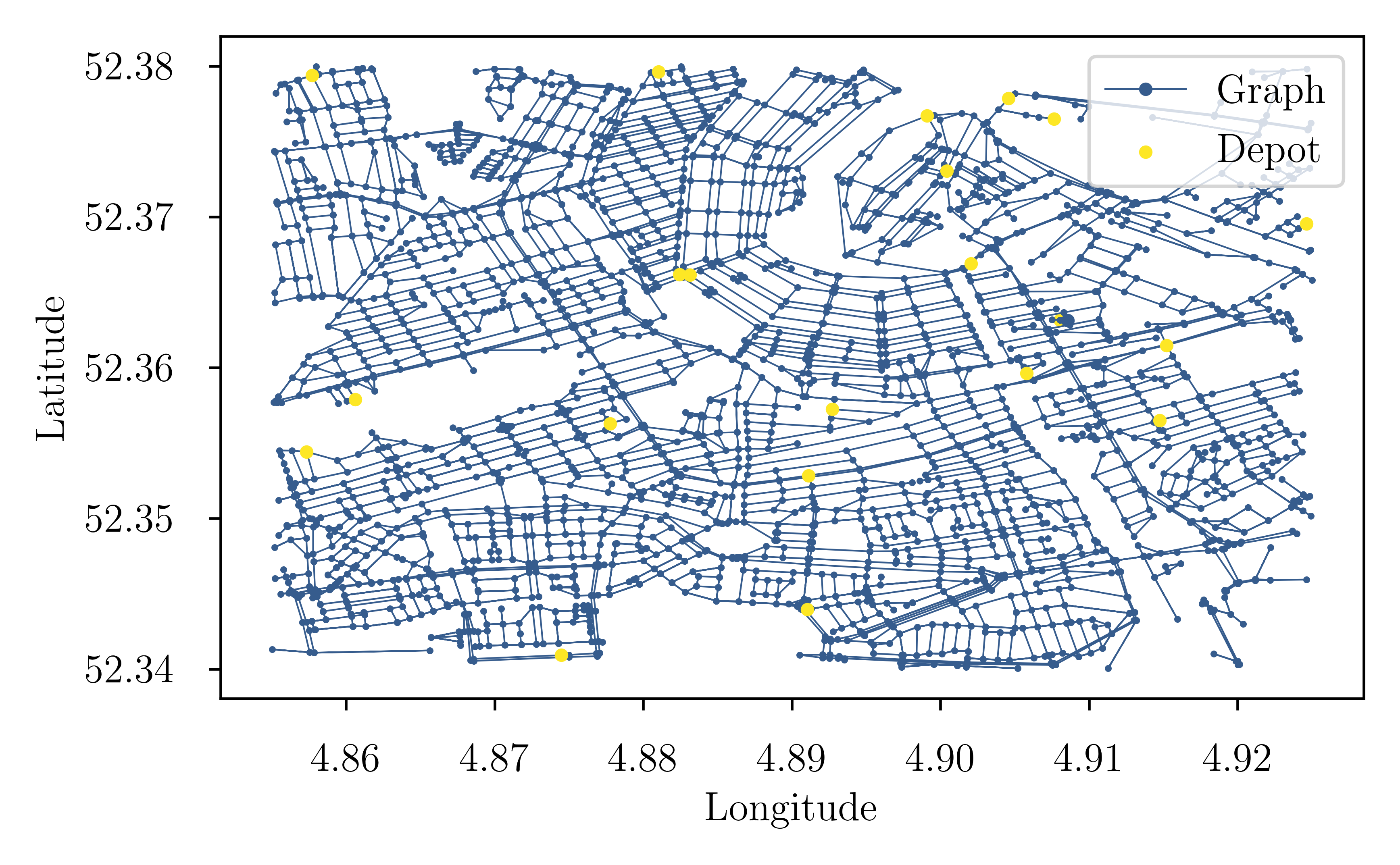}}}\hspace{5pt}
\subfloat[Temporal distribution of placed orders.]{%
\resizebox*{7cm}{!}{\includegraphics[width=1.\textwidth]{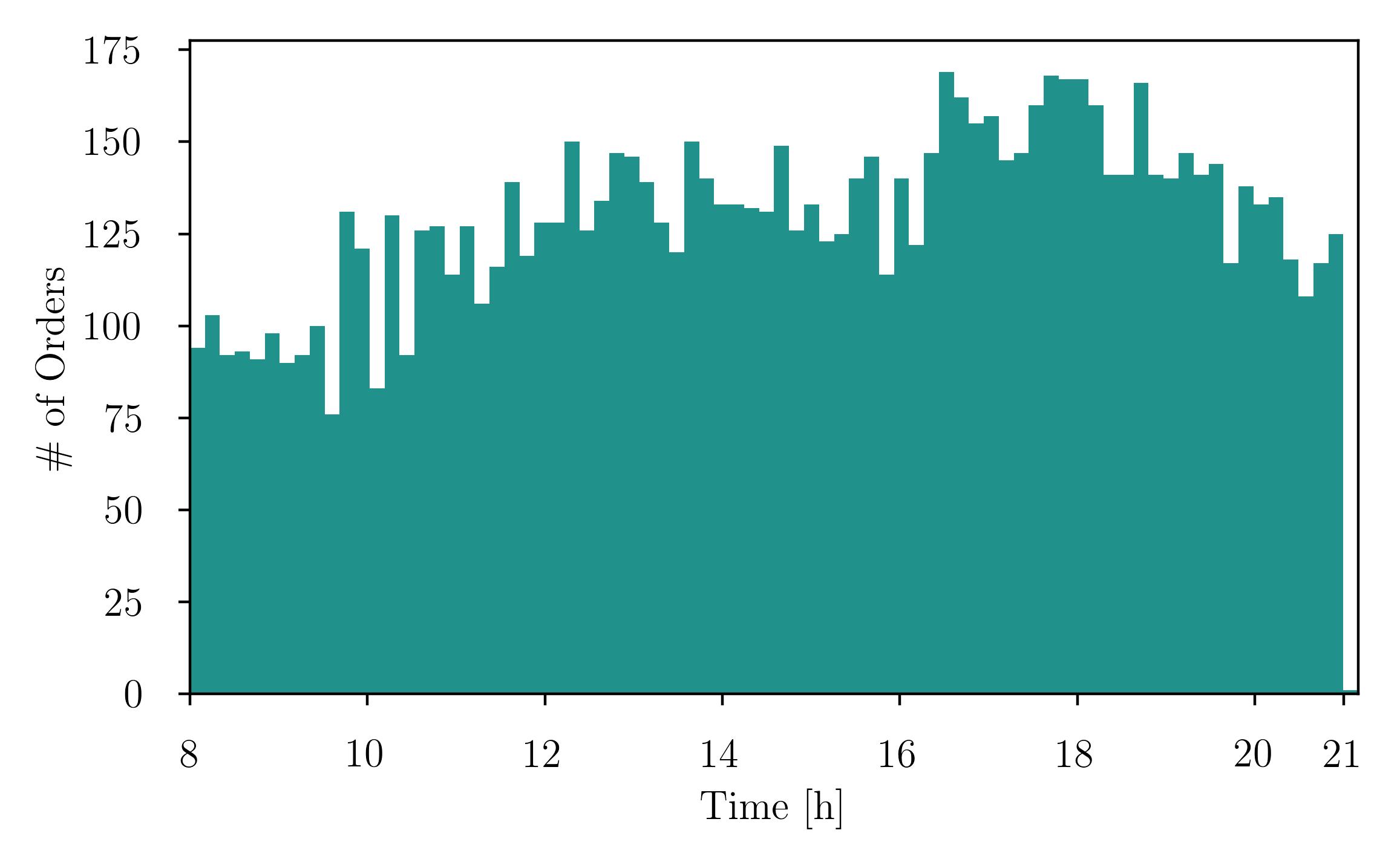}}}
\caption{A visual representation of the underlying graph $G=(\mathcal{N},\mathcal{A})$ is shown on the left. The locations of all 20 depots are highlighted in yellow. On the right side, the temporal distribution of all order's request times $t_o \,\, \forall o \in \mathcal{O}$ is depicted. Each bar shows the number of newly placed orders within ten minutes.} \label{fig:graph_and_demand}
\end{figure}

To analyze the proposed algorithm, we simulate a potential day in the city centre of Amsterdam. We represent the street network as a directed graph containing 2717 nodes and 5632 edges, shown in Figure \ref{fig:graph_and_demand}(a). Over the whole service area, 20 depots to pick up orders are distributed following a k-center algorithm, which is explained in the Appendix \ref{ap:k-center}. The travel times between nodes are calculated as their distance divided by a constant speed of $36 \frac{\text{km}}{\text{h}}$. The simulated demand pattern features 10,000 orders, homogeneously distributed in space, covering a period from 08:00 to 21:10, including a noon and a stronger evening peak. The according distribution in time is shown in Figure \ref{fig:graph_and_demand}(b). Each bar shows the number of newly placed orders within ten minutes. In the last 10 minutes, before the end of the day $T_{end}$, no more orders are placed, $\delta_T =10$.

30 autonomous vehicles with a maximum capacity of six are used. The maximum trip size $\eta$ is set to ten. The maximum delay $\delta_{\text{delay,real}}$ is set as eight minutes and equal to $\delta_{\text{delay,heuristic}}$, resulting in a $\zeta$ of one. Per order, the three closest depots to the final destination ($x=3$) are considered. To load and service an order, we assume $\delta_{load}=15 \sec$, implying that all orders are prepared in advance and only need to be loaded, and $\delta_{service} = 30 \sec$, assuming that all customers are ready to grab their groceries at the front door. The algorithm runs in time spans $\Delta t=$ 100 seconds. The penalty for ignoring an order is set to equal $10^4$ seconds. We weighted the two different objectives with $\beta=1/3$. These values have been chosen to create a scenario serving most orders with limited resources such that the system is forced to work as efficiently as possible but can't serve everything. To solve the ILP described in Algorithm \ref{alg:opt}, we use the software Mosek 7.1 with a time-budget of 50 seconds. This time-budget is enough to find the optimal solutions in about 85\% of the cases, otherwise the best obtained solution at that point is used. All parameters and their respective values are summarized in Table \ref{tb:para_base} in the Appendix. 
As the problem formulation considers several conflicting objectives, we report all the relevant operational measures. Note that the value of the objective function is not very informative, as the rejection penalty $\alpha$ dominates.\\

\begin{figure}
    \centering
	\includegraphics[width=0.5\textwidth]{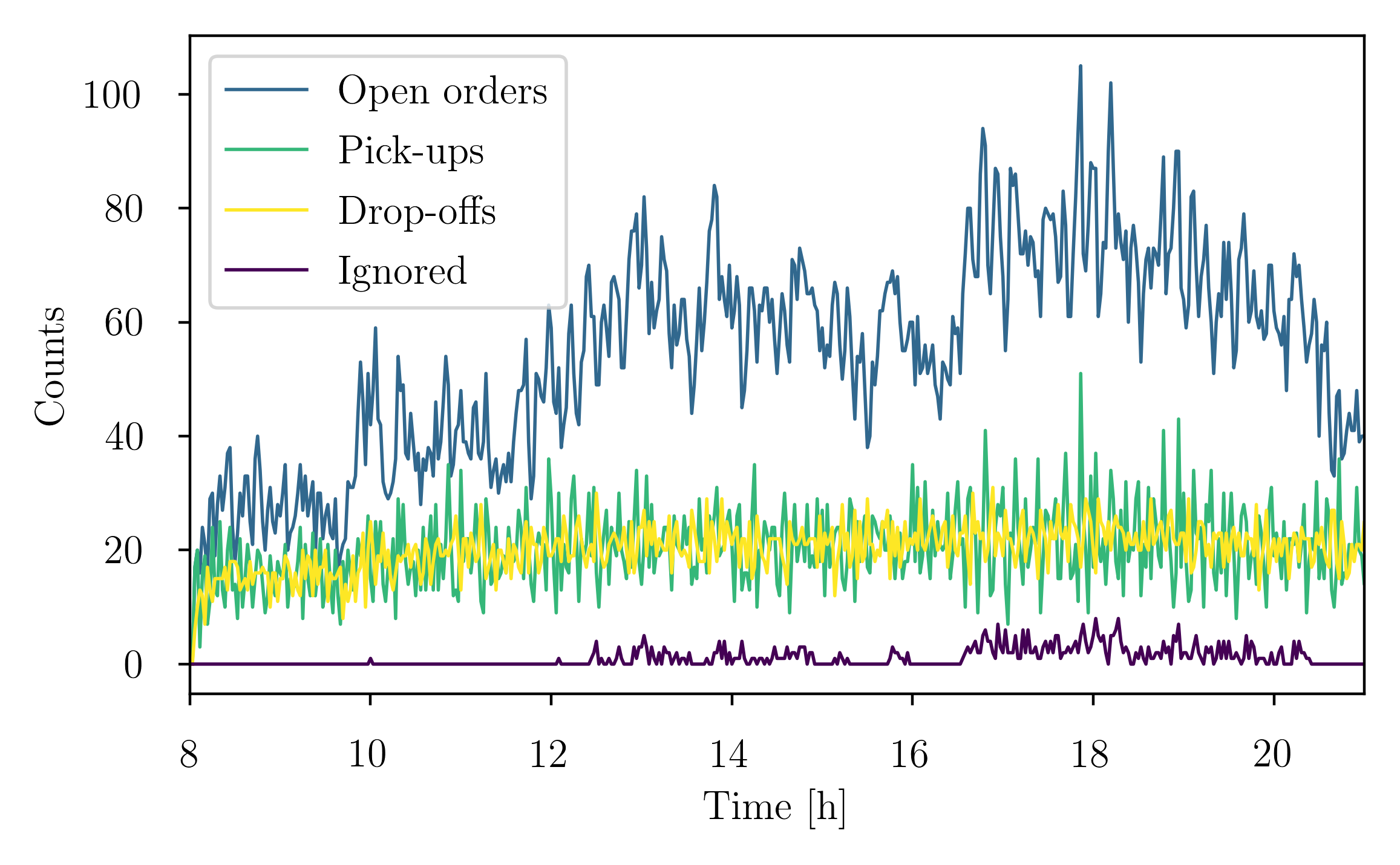}       
	\caption{Different actions and number of open orders over time, counts are evaluated per time step.}
	\label{fig:stati_parcles}
\end{figure}

\begin{figure}
\centering
{%
\resizebox*{3cm}{!}{\includegraphics[width=1.\textwidth]{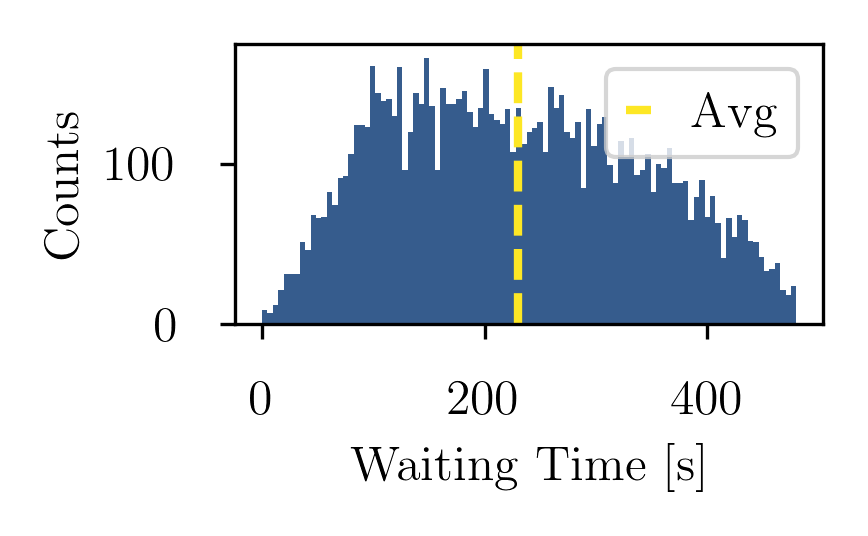}}}\hspace{5pt}
{%
\resizebox*{3cm}{!}{\includegraphics[width=1.\textwidth]{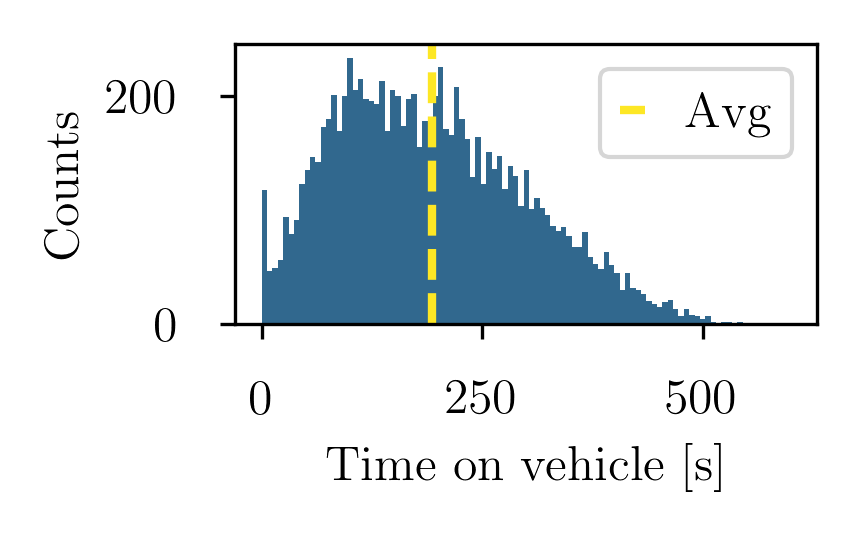}}}\hspace{5pt}
{%
\resizebox*{3cm}{!}{\includegraphics[width=1.\textwidth]{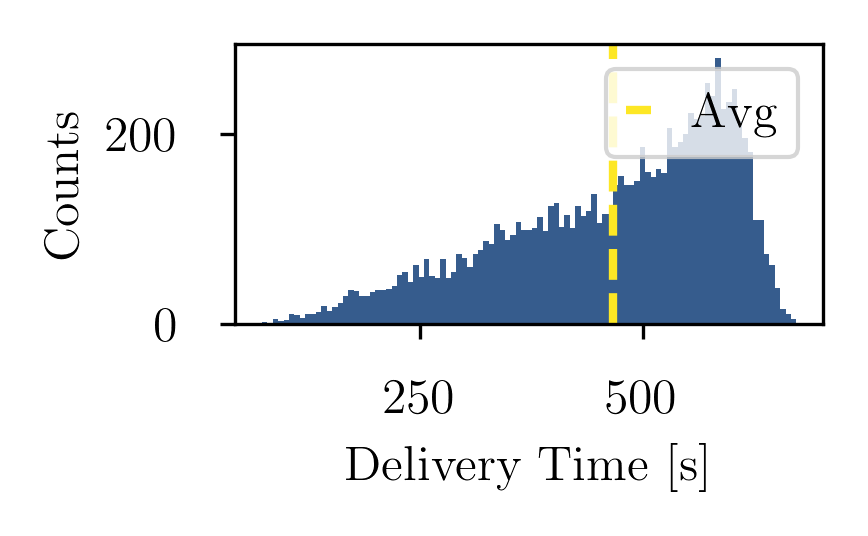}}}\hspace{5pt}
{%
\resizebox*{3cm}{!}{\includegraphics[width=1.\textwidth]{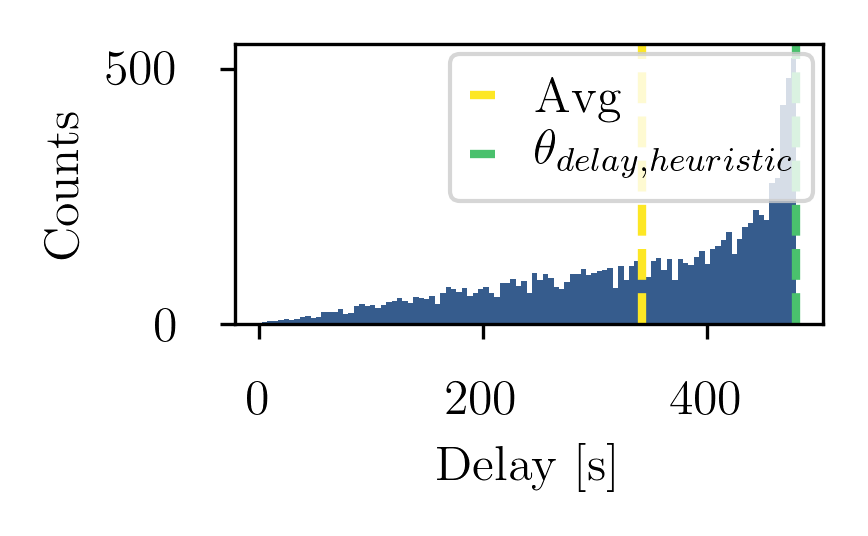}}}
\caption{Distributions of time to pick-up, time of parcels spend loaded to a vehicle, the total delivery time, and delay of the base scenario.} \label{fig:time_dis}
\end{figure}

First, we evaluate the service rate, which is defined by the percentage of served orders in respect to all requested orders. A service rate of $95.19\;\%$ is achieved, which equals 481 ignored orders. Figure \ref{fig:stati_parcles} shows the number of open orders, pick-ups and drop-offs and the number of finally ignored orders per time step $\Delta t$. Most ignored orders happen during peak times. Peak times are characterized by the high workload, represented in the high number of open orders.
The number of pick-ups shows occasional spikes. These appear as vehicles can load a high number of parcels consecutively without driving when they visit a depot.
Second, we analyze different time spans (time KPIs) involved in the delivery process of each order, see Figure \ref{fig:times}. The distributions of the time until pick-up (mean:\;3\,min\,50\,s), the time a parcel is loaded onto a vehicle (mean:\;3\,min\,13\,s), the delivery time (mean:\;7\,min\,47\,s), and the associated delay (mean:\;5\,min\,43\,s) are illustrated in Figure \ref{fig:time_dis}. These times can be compared to the average distance of all nodes to their closest depot, which is is 1\,min\,20\,s. Note that the total delivery time is always greater than 45s, the sum of the loading and service time ($\delta_{load} + \delta_{service}$). The delay distribution increases strongly towards a sharp cut-off at 480s/8min, the maximum allowed delay.\\

\begin{figure}
\centering
\subfloat[Delay distribution 9a.m. - 11a.m.]{%
\resizebox*{7cm}{!}{\includegraphics[width=1.\textwidth]{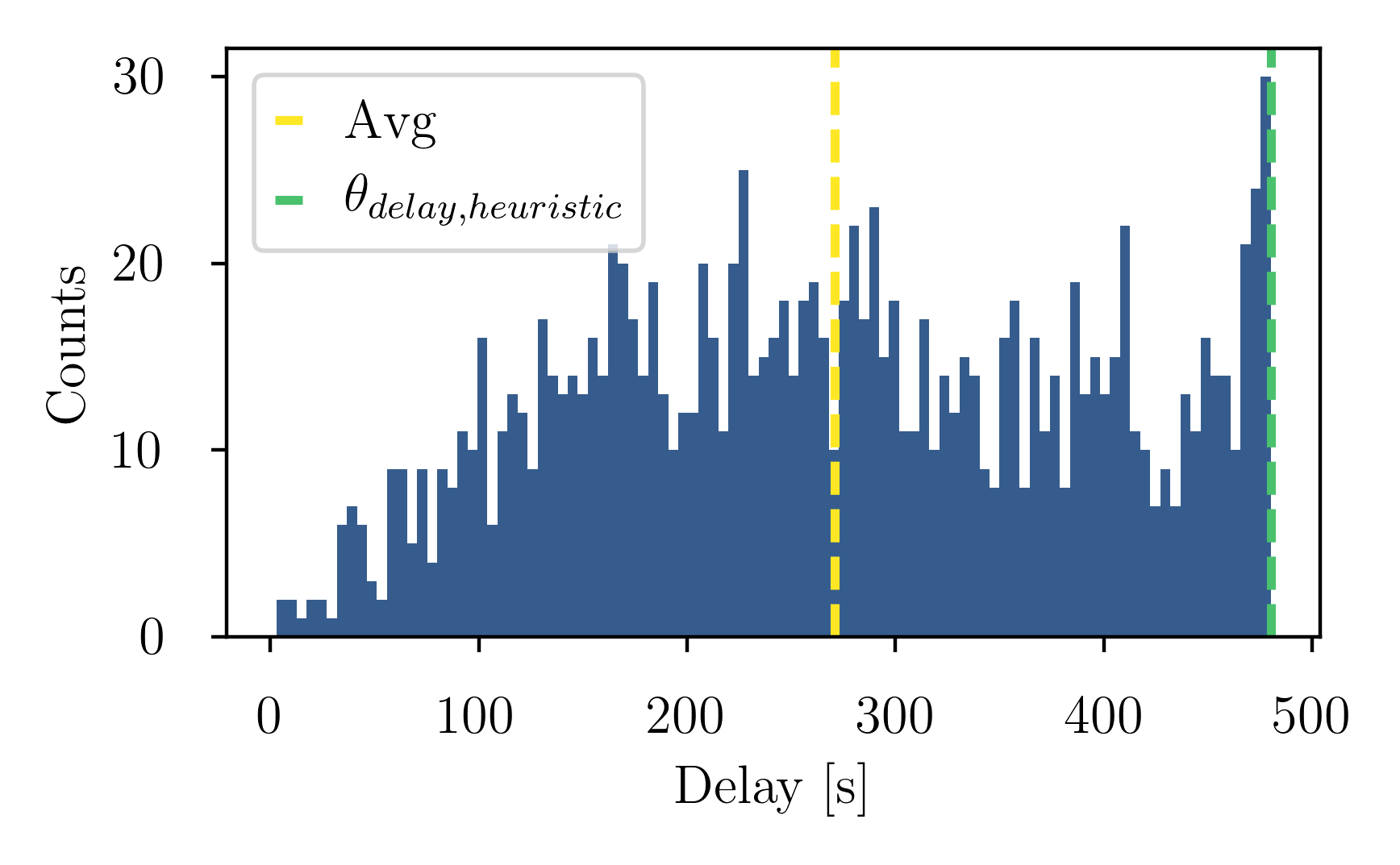}}}\hspace{5pt}
\subfloat[Delay distribution 5p.m. - 7p.m.]{%
\resizebox*{7cm}{!}{\includegraphics[width=1.\textwidth]{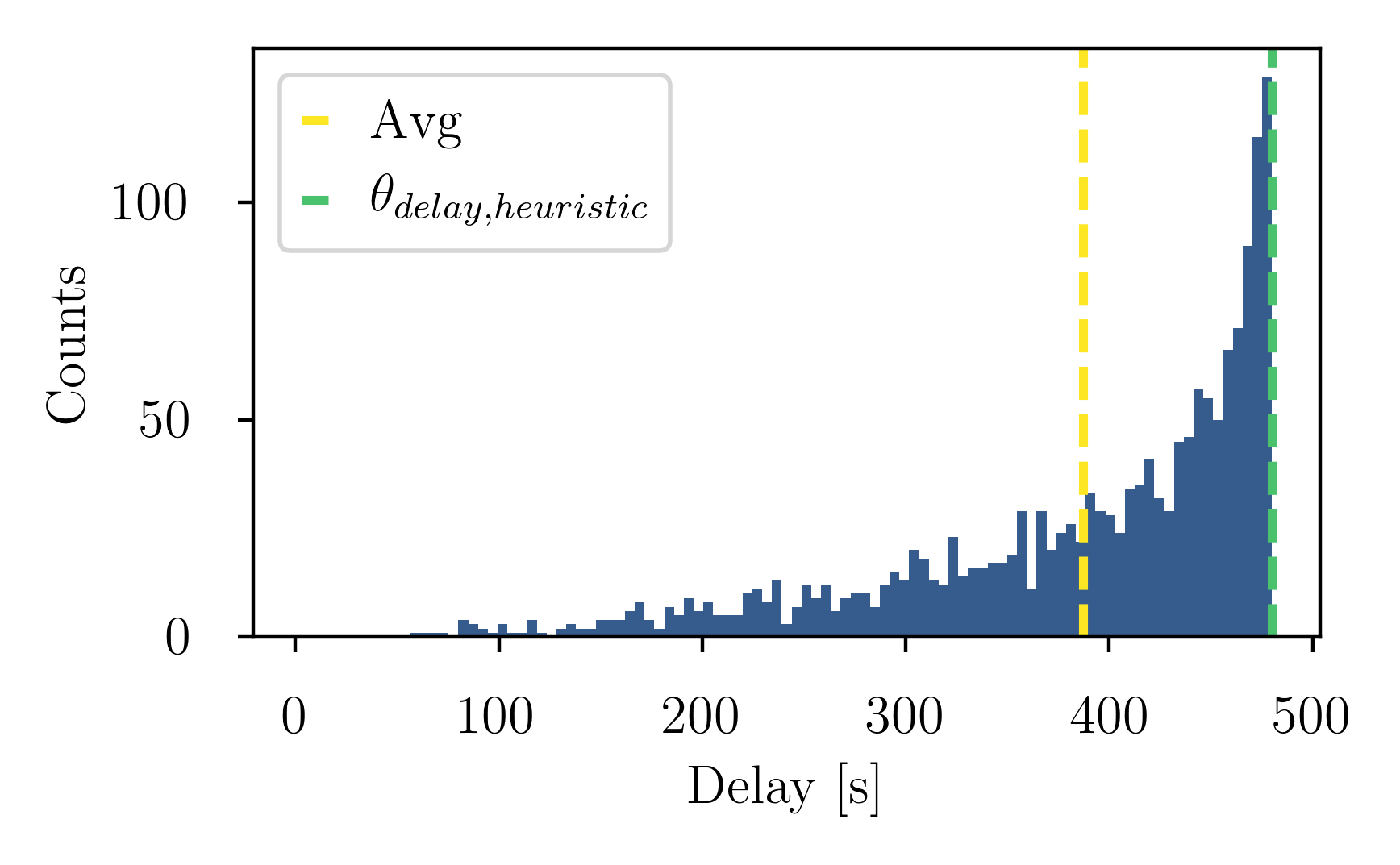}}}
\caption{Distribution of delay for two different time windows differing in workload of the base scenario.} \label{fig:delay_temporal}
\end{figure}

Let us analyze the delay in more detail. We distinguish two time windows of two hours, one in the morning (09:00 to 11:00) with low workload and one in the evening (17:00 to 19:00) with a high workload. Figure \ref{fig:delay_temporal} shows the delay distribution for all orders placed in the corresponding time windows. For low workload (Figure \ref{fig:delay_temporal}(a)), the average delay is significantly lower and the overall shape of the distribution is less pushed towards the maximum delay. As low workload results in more free resources, the service level can be improved and the resources are not needed to serve more orders in the first place. This is also reflected in the number of rejected orders, as shown in Figure \ref{fig:stati_parcles}. In contrast, during high workload (Figure \ref{fig:delay_temporal}(b)) most orders are served with a high delay.

\begin{figure}
\centering
\subfloat[Occupancy.]{%
\resizebox*{7cm}{!}{\includegraphics[width=1.\textwidth]{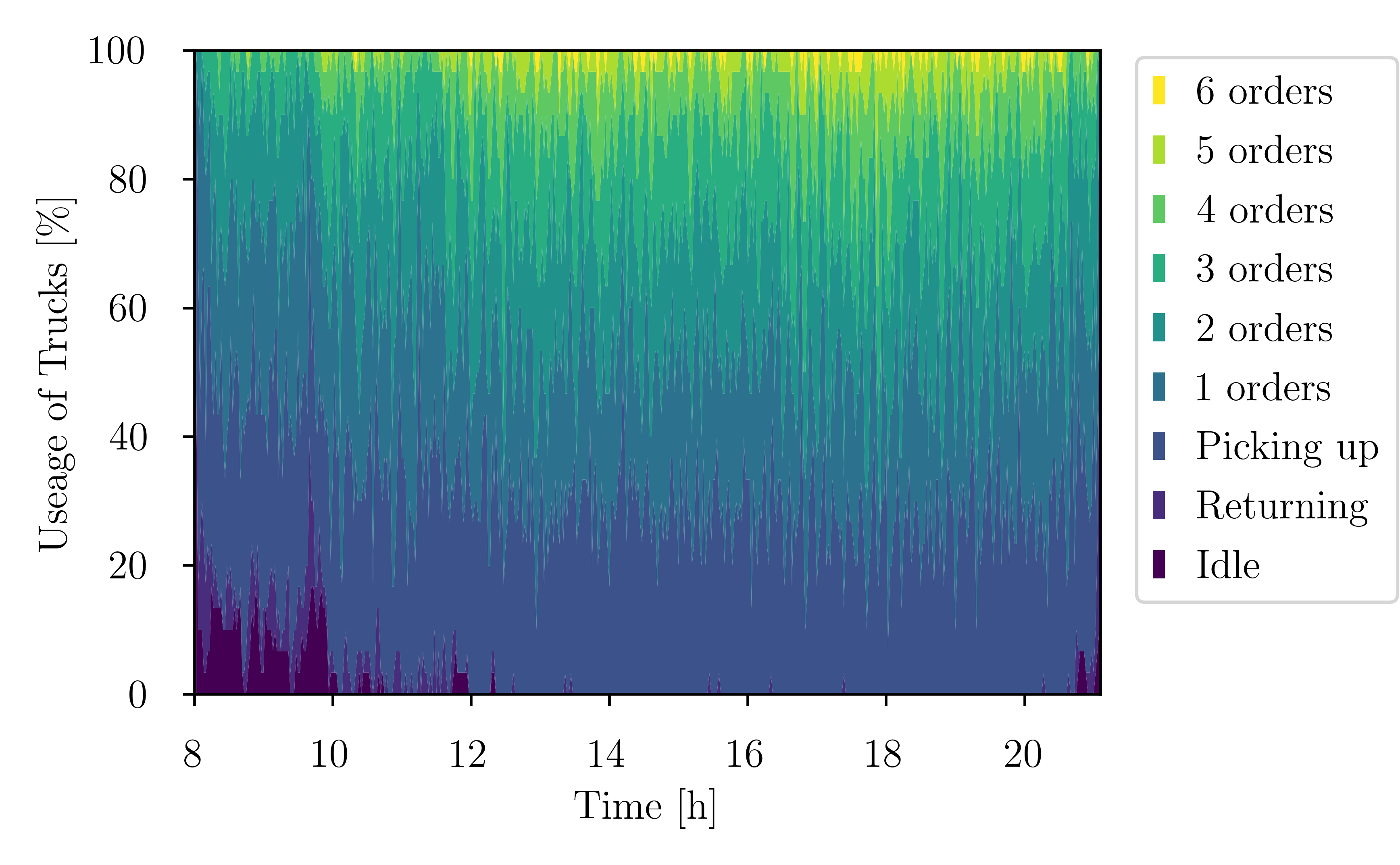}}}\hspace{5pt}
\subfloat[Mean loaded parcels.]{%
\resizebox*{7cm}{!}{\includegraphics[width=1.\textwidth]{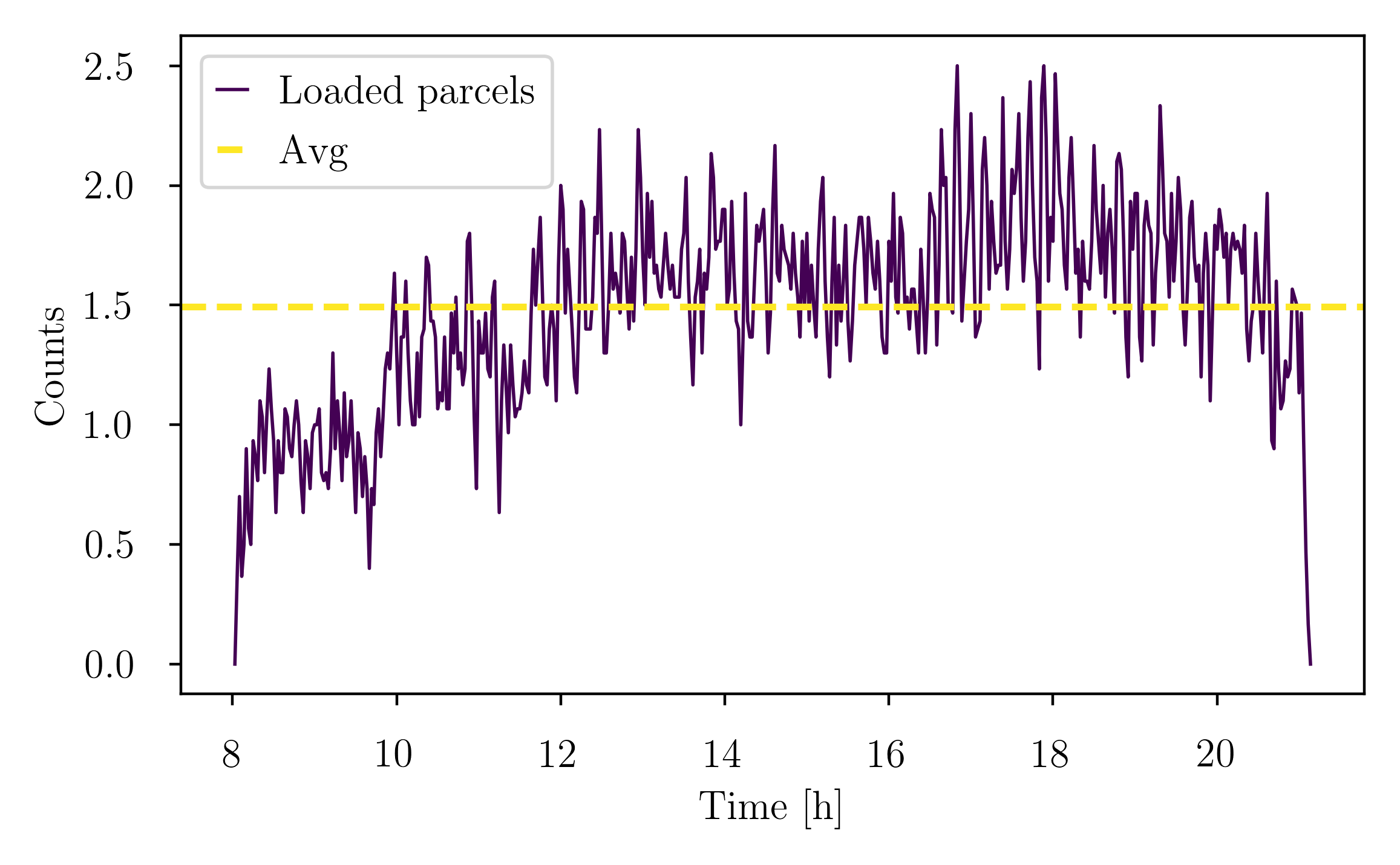}}}
\caption{On the left, occupancy of all vehicles over the entire operation duration, the evening peak can be identified clearly. On the right, the mean number of loaded parcels of all vehicles as the day progresses.} \label{fig:vehicle_use}
\end{figure}

Third, we analyze how the proposed method utilizes each vehicle. The occupancy of all vehicles is depicted in Figure \ref{fig:vehicle_use}(a). The evening peak of the demand can also be identified through the brighter colors that appear there, meaning that many vehicles have more loaded parcels. On the other hand, idle vehicles only occur at the beginning and end of the day. During the rest of the day, vehicles are immediately used while or after returning to a depot. A summary of Figure \ref{fig:vehicle_use}(a) is shown in Figure \ref{fig:vehicle_use}(b), displaying the mean number of loaded parcels of all vehicles over time. The average load per vehicle over the day is 1.49 parcels.\\

Fourth, we analyze the total travelled distance. In the base scenario, a distance of 8,973.8\,km is travelled by all 30 vehicles. All vehicles are used similarly for the whole day. Driven distance per vehicle ranges from 267.86\,km to 311.86\,km.\\
%%%%%%%%%%%%%%%%%%%% END BASELINE %%%%%%%%%%%%%%%%%%%%%%%%%%%%%%%%%%%%%%%%%

\subsection{Comparison} \label{sec:result_compare}
We now assess the merits of our method by comparing it to different approaches. First, we compare it with the results that would be obtained by using a greedy assignment strategy (Section \ref{sec:result_greedy}). Moreover, we analyze the merits of the main features allowed by our method, namely considering multiple depots (Section \ref{sec:results_compare_single}), and the benefits of pre-empty depot returns (Section \ref{sec:results_compare_no_preempty}).
For simplicity, we refer to the previously analyzed scenario as the base scenario (Section \ref{sec:result_base}). All parameters are fixed to the same values as in the base scenario (Table \ref{tb:para_base}) until mentioned otherwise.

\subsubsection{Greedy Assignment Strategy} \label{sec:result_greedy}
\begin{figure}[H]
	\centering
	\includegraphics[width=1.0\textwidth]{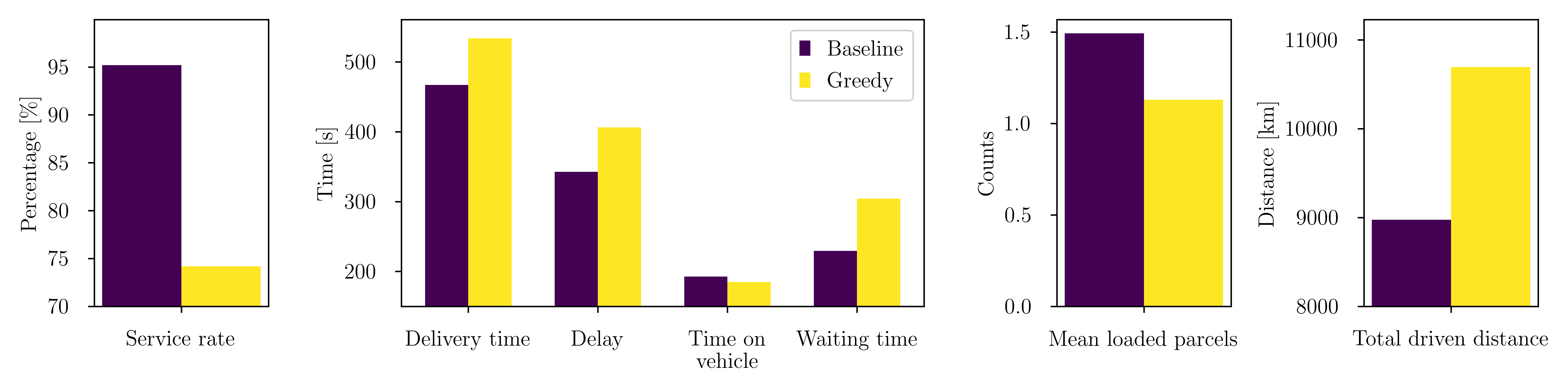}
	\caption{Service rate, time KPIs, mean loaded parcels and total driven distance of the base scenario and the greedy assignment strategy.}
	\label{fig:compaire_greedy}
\end{figure}

This section compares the base scenario to the following greedy assignment strategy.\\
\textbf{The Greedy Strategy:} Every time an order gets placed, we check immediately which is the best way to serve it. To do this, we check how the new order could be inserted into each vehicle. Then, the new order is assigned to the best vehicle, i.e., the one that would achieve the minimum added cost if inserting the new order (all $x$ pick-up options are considered) into the current plan of the vehicle. Cost is measured according to the used objective function (Equation \ref{eq:gamma}). If an order can not be added to any vehicle's route without violating some constraint, it is rejected immediately. This strategy is presented in Algorithm \ref{alg:greedy_full} using a function called $FindBestVehicle(o,\mathcal{V})$. It either outputs the best vehicle $v \in \mathcal{V}$ to serve the order $o$, or a statement that there is no feasible way to insert the new order within any vehicle without violating any constraint.\\
The obtained results are visualized in Figure \ref{fig:compaire_greedy}. We see a decrease of the service rate from  95.19\% to 74.18\%, while delay and total driven distance increase significantly. This means that with the greedy algorithm fewer orders are delivered, those that are delivered require longer travel times, and total operators costs become larger.\\

\begin{algorithm}[h]
	\caption{Greedy Assignment Strategy}
	\label{alg:greedy_full}
	\SetKwInOut{Input}{input}
	\SetKwInOut{Output}{output}
	\Input{Demand $\mathcal{O}$, Vehicle fleet $\mathcal{V}$, Graph $G$, Depots $\mathcal{D}$}
	\Output{Greedy solution to the overall problem $\Omega_{\text{greedy}}$}
	\SetKwBlock{Beginn}{beginn}{ende}
	\Begin{
		\For{$ o \in \mathcal{O} \, 
		\text{\scriptsize (orders are sorted by their request times)}$}{
		    $t = t_o$\; 
			$v =FindBestVehicle(o,\mathcal{V}_t)$\;
			\If{vehicle found $\text{\scriptsize  (order can be assigned to a vehicle without violating any constraint)}$}{
			    $\Omega_{\text{greedy}} \leftarrow v,o \, \text{\scriptsize (assign order $o$ and vehicle $v$)}$ \;}
			\Else{
		        $\mathcal{IO} \leftarrow o \, \text{\scriptsize   (no feasible vehicle thus order gets rejected)}$\;}
		    Propagate time to next order\;
		    Update vehicle fleet $\mathcal{V}_t$\;
		}%\EndFor
		\Return $\Omega_{\text{greedy}}$
	}
\end{algorithm} 

\subsubsection{Considering the closest depot only} \label{sec:results_compare_single}
\begin{figure}[H]
	\centering
	\includegraphics[width=1.0\textwidth]{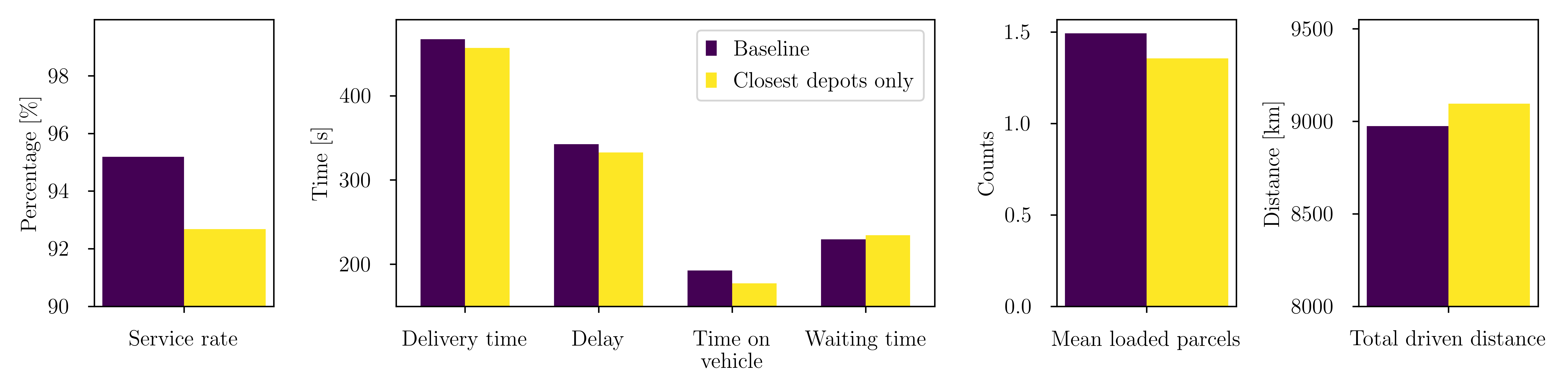}
	\caption{Service rate, time KPIs, mean loaded parcels and total driven distance of the base scenario and the same scenario considering the closest depot only.}
	\label{fig:compaire_singleDepot}
\end{figure}

We compare the base scenario to the case in which each order is picked up at the depot that is closest to its destination (i.e., $x=1$). This is similar to comparing to the case in which the problem is decomposed into several single-depot problems, although this approach is still more flexible as vehicles are not fixed to some specific depot. Hence, the results we obtain would be even stronger if compared to a decomposition approach.\\
For $x=1$, worse results in service rate and total driven distance are obtained, as shown in Figure \ref{fig:compaire_singleDepot}. The service rate drops from 95.19\% to 92.68\% i.e., 251 additionally ignored orders. Even having fewer parcels delivered overall, the total driven distance increases by 121.14\,km, for $x=1$. These improvements of the baseline come at the cost of increased delay in the order of seconds, it increases from 5\,min\,33\,s to 5\,min\,43\,s. See Section \ref{subsec:influence_x} for the analysis of additional values of $x$.

\subsubsection{No pre-empty depot returns} \label{sec:results_compare_no_preempty}
\begin{figure}[H]
	\centering
	\includegraphics[width=1.0\textwidth]{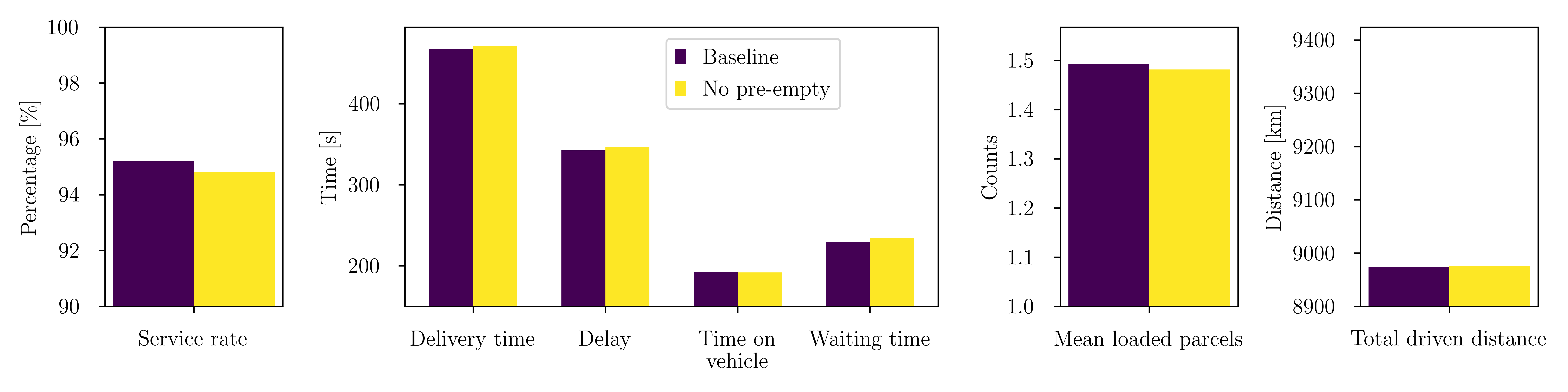}
	\caption{Service rate, time KPIs, mean loaded parcels and total driven distance of the base scenario and the same scenario with prohibiting pre-empty depot returns.}
	\label{fig:compaire_preempty}
\end{figure}

One of the virtues of our approach is that it allows for pre-empty depot returns. We now compare to the case where we prohibit those depot returns, meaning that only empty vehicles can load new orders. Results are depicted in Figure \ref{fig:compaire_preempty}, and show a decrease in service rate (95.19\% $\rightarrow$ 94.81\%) and a slight increase for all time KPIs (average delay: 5\,min\,43\,s $\rightarrow$ 5\,min\,46\,s) and also for the total driven distance (8,973.8\,km $\rightarrow$ 8,975.5\,km). 
Generally, the more depots are located within a service area the less impactful allowing for pre-empty depot returns is (i.e., changes would be more significant for a total of 5 or 10 depots.) This is due to a smaller average distance of vehicles to the next depot. We remark that these improvements are fully achieved by modifying the routes without the need for additional infrastructure.

\subsection{Sensitivity Analysis}  \label{sec:result_sensitivity}
To study the effect of specific parameters of our proposed method, we systematically modify them while fixing all others (Table \ref{tb:para_base}). We dedicate one subsection to each studied parameter. All unchanged parameters are chosen as in the base scenario, analyzed in Section \ref{sec:result_base}. 

\subsubsection{Number of considered depots per order, $x$} \label{subsec:influence_x}
\begin{figure}[H]
	\centering
	\includegraphics[width=1.0\textwidth]{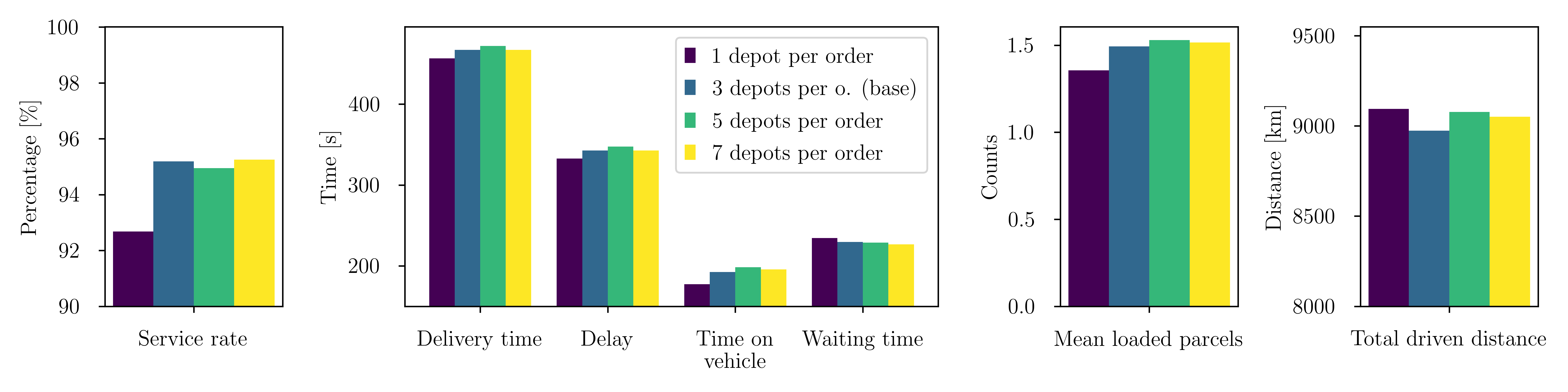}
	\caption{Service rate, time KPIs, mean loaded parcels, and total driven distance of the different runs, featuring a different number of considered depots as potential pick-up locations per order $x$.}
	\label{fig:kpi_vs_consideredDepots}
\end{figure}

\begin{figure}
\centering
\subfloat[$x=3$ (Base scenario).]{%
\resizebox*{7cm}{!}{\includegraphics[width=1.\textwidth]{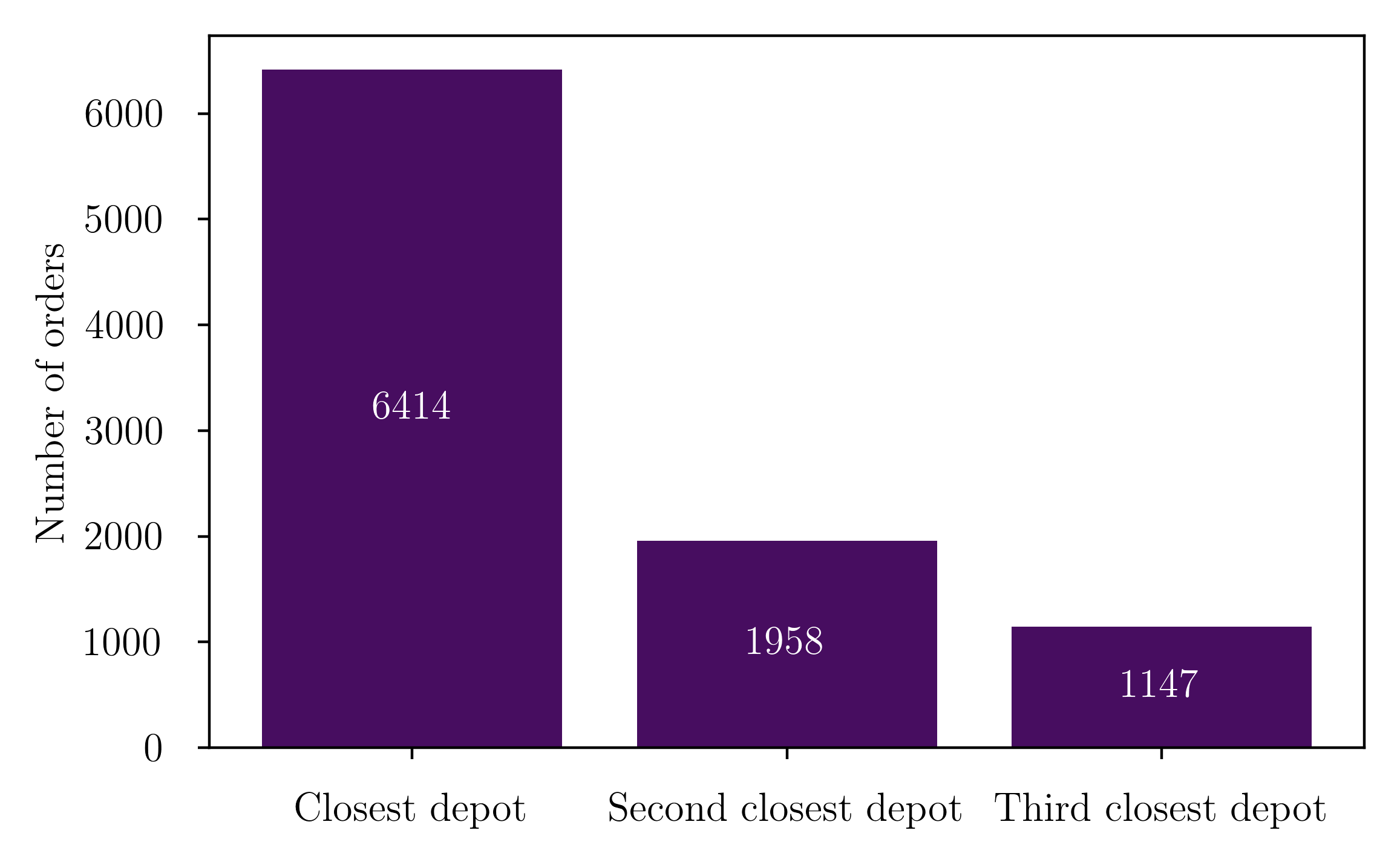}}}\hspace{5pt}
\subfloat[$x=5$.]{%
\resizebox*{7cm}{!}{\includegraphics[width=1.\textwidth]{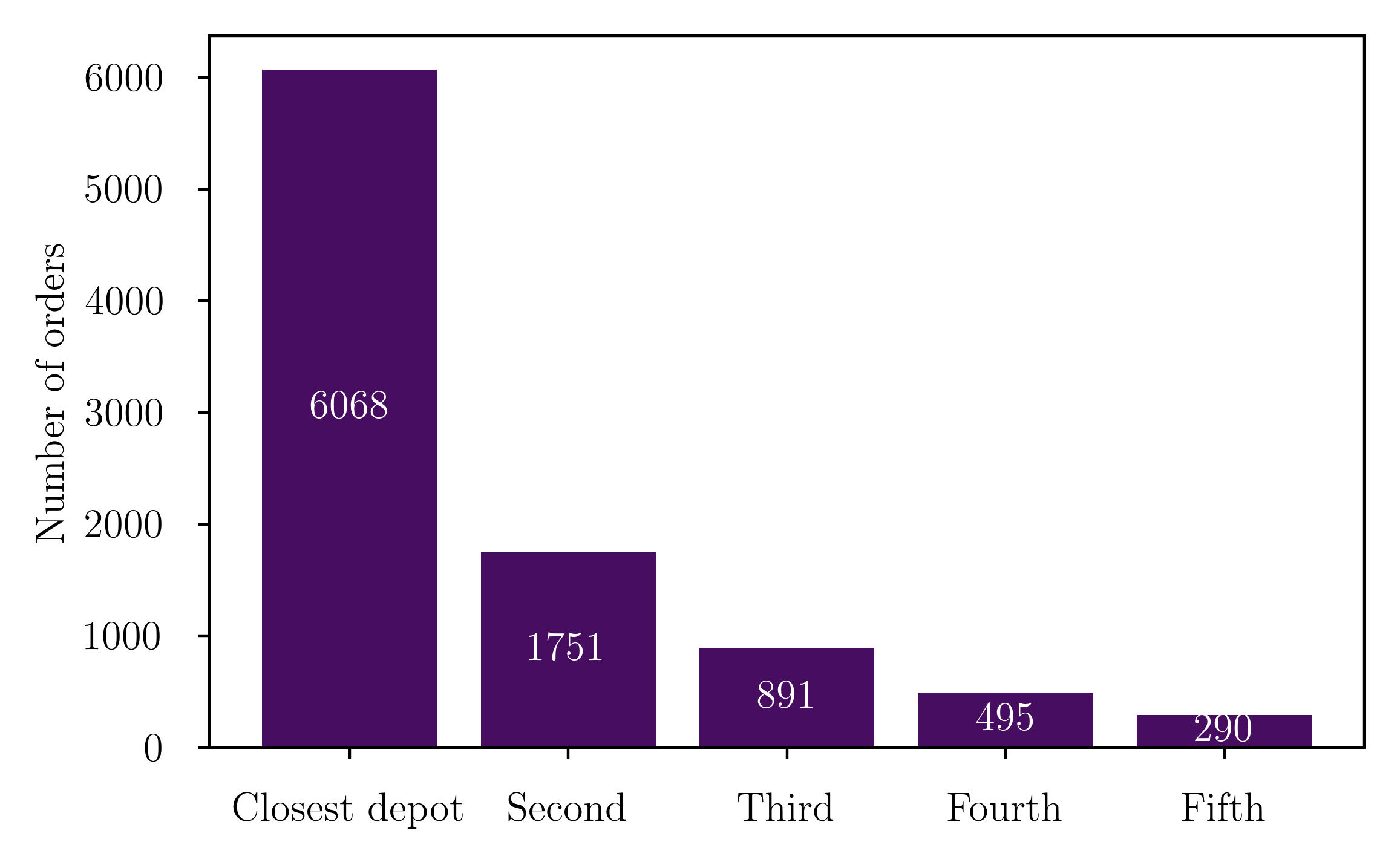}}}
\caption{The usage of depots, ranked according to the distance to their order's destination, for the baseline (left) and if five depots are considered per order (right).} \label{fig:depot_useage}
\end{figure}

In Section \ref{sec:method_pickup} we introduced a heuristic, namely that only the $x$ closest depots to an order's destination are considered as potential pick-up locations. The more depots are considered the higher the number of candidates will be, yet, the higher the computational burden. Figure \ref{fig:kpi_vs_consideredDepots} shows the results if one (see \ref{sec:results_compare_single}), three, five, or seven depots per order are considered.
Service rate rises while total travelled distance decreases when three instead of one depot per order are considered. Both get worse if $x=5$ and then improve slightly if $x=7$. Delay increases the more depots are considered with a slight drop if $x$ changes from five to seven.
Figure \ref{fig:depot_useage} additionally shows the usage of depots, if they are ranked according to the distance to their order's destination, for the baseline (Figure \ref{fig:depot_useage}(a)) and the scenario considering five depots per order (Figure \ref{fig:depot_useage}(b)). In both cases, the closer the depot, the more frequently it is used, with the most significant change from the closest to the second closest depot. The closest depot was used in both cases over 6000 times, leaving over 30\% to be distributed over the others, showing again that it is beneficial to consider them. 

On the one hand, considering multiple depots is beneficial, as shown by the improvements compared with considering the closest depot only. On the other hand, it is not always better to consider more depots per order, as revealed by the worse results obtained with $x=5$. This can be explained by the myopic nature of our approach, which can lead the system into unfavourable states to serve future demand. For a single state (i.e. for one assignment decision), using more depots is better as it enlarges the set of feasible solutions, but the overall problem's solution can worsen due to chaining multiple states dynamically with each other. For example, a truck might be sent to a depot that is far away when considering only current information, but if some orders appear nearby soon after, it would have been better to go to a closer depot in the first place. Thus, those results confirm the benefits of introducing a heuristic. We conclude: 
\begin{itemize}
	\item Considering only one depot per oder, as \cite{yu_improved_2013} and \cite{xu_hybrid_2018}, can be inferior to considering multiple ones. 
	\item To consider as many depots as possible can be inefficient for dynamic problems having imperfect anticipation.
\end{itemize}
The question "How many depots should be considered?" emerges. The answer can depend on various factors, including the problem at hand or even the current state. This is outside of the scope of this work and is left for future work.\\

\subsubsection{Total number of depots $H$} \label{sec:Results_tot_depot}
\begin{figure}[H]
	\centering
	\includegraphics[width=1.0\textwidth]{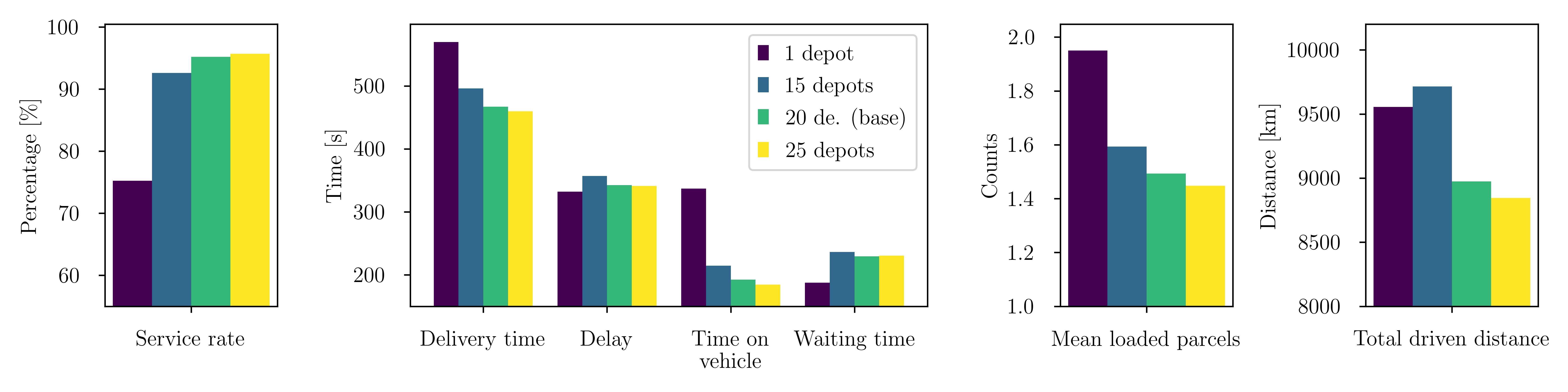}
	\caption{Service rate, time KPIs, mean loaded parcels, and total driven distance of the four different runs, featuring a different number of total depots distributed over the service area.}
	\label{fig:kpi_vs_totStores}
\end{figure}

We varied the number of placed depots within the service area. We simulated scenarios featuring 1, 15, and 25 depots in total. Results are depicted in Figure \ref{fig:kpi_vs_totStores}.
Service rate improves at decreasing rates the more depots are available.
Delay shows a non-monotonic behaviour, which is related to the corresponding service rates. For a single depot, fewer orders are served, i.e., more orders are ignored. When some orders are ignored, the most complicated ones are ignored first, meaning that they would have had a high delay if delivered.
The number of mean loaded parcels decreases the more depots are available. Total driven distance generally decreases for more depots, except for 15 depots compared to the single depot case, explainable by the substantial increase in service rate. Generally, the magnitude of changes in performance varies while the number of depots is increased linearly in steps of five.
These results suggest the question: "How many depots are optimal, taking the cost to open and operate them into account?". We consider this question interesting and relevant for future research related to the multi-facility location problem (\cite{farahani2009facility}).\\

\subsubsection{Allowing for reinsertion of orders}
\begin{figure}[H]
	\centering
	\includegraphics[width=1.0\textwidth]{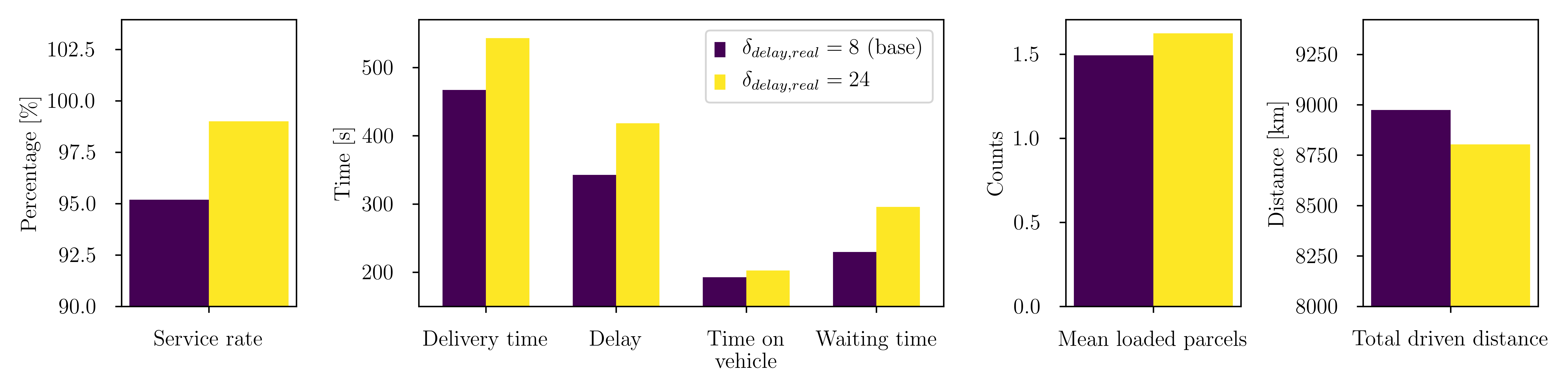}
	\caption{Service rate, time KPIs, mean loaded parcels, and total driven distance of the base scenario and the same scenario having a $\delta_{\text{delay,real}}$ of 24 minutes, thus allowing reinsertion.}
	\label{fig:sensitvity_reinsert}
\end{figure}

\begin{figure}[H]
\centering
\subfloat[Delay distribution.]{%
\resizebox*{7cm}{!}{\includegraphics[width=1.\textwidth]{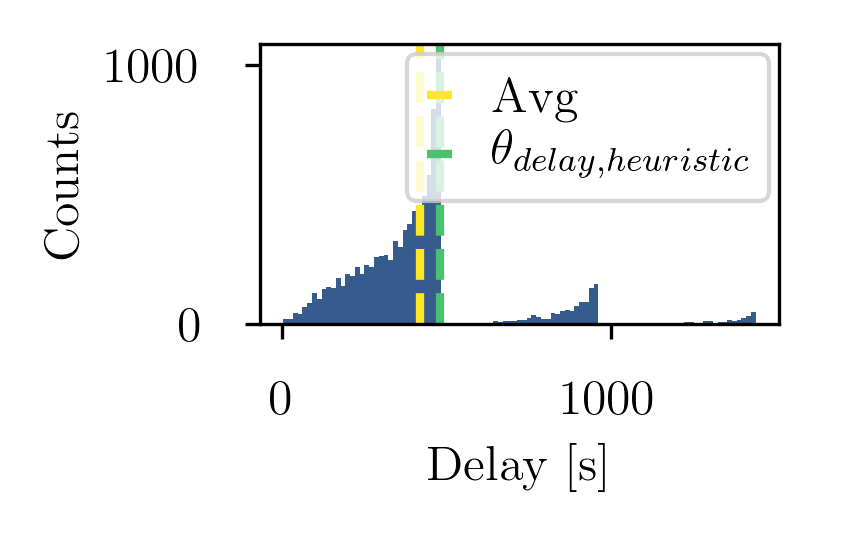}}}\hspace{5pt}
\subfloat[Reinsertion count.]{%
\resizebox*{7cm}{!}{\includegraphics[width=1.\textwidth]{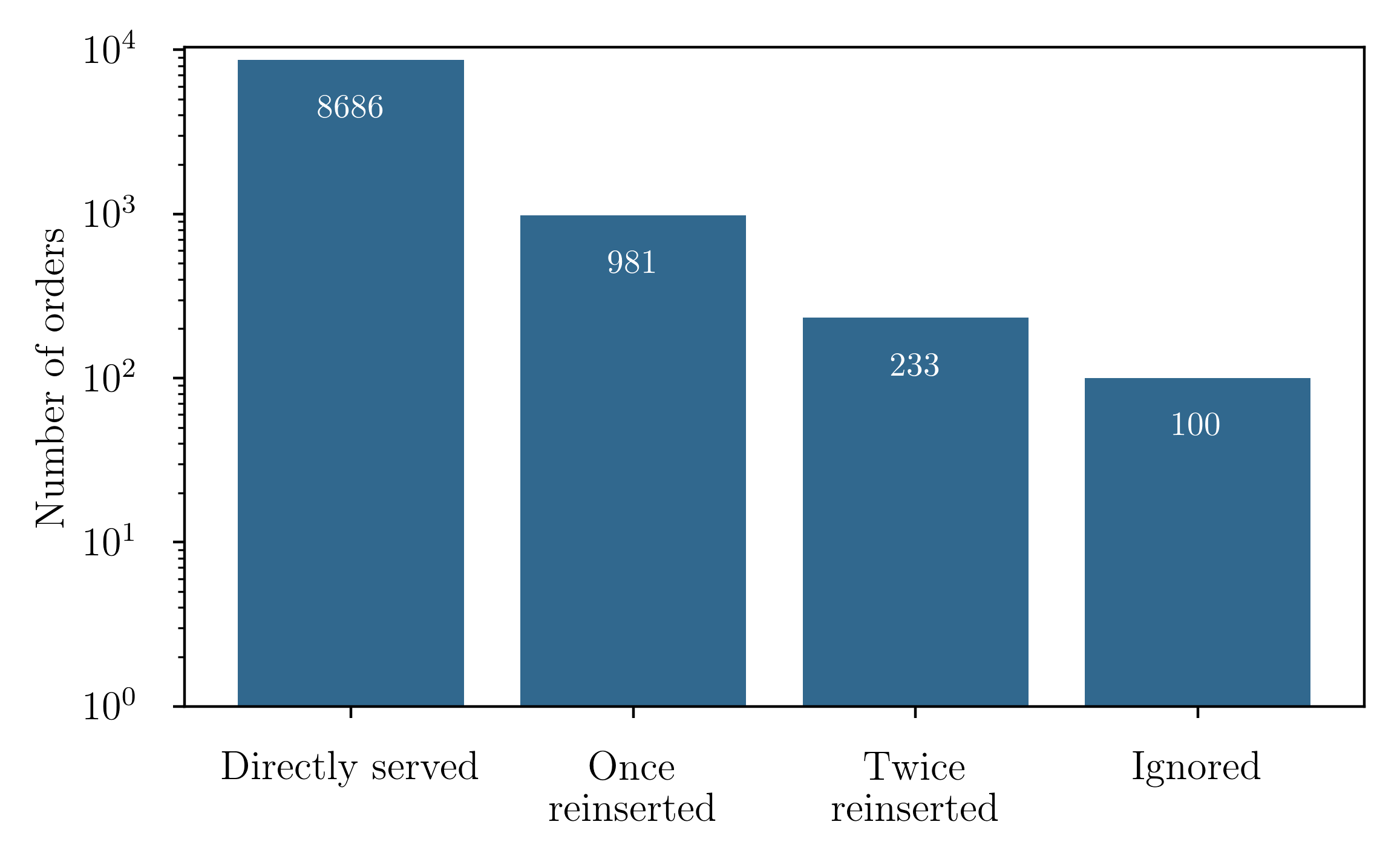}}}
\caption{The delay distribution having a $\delta_{\text{delay,real}}$ of 24 minutes, thus allowing reinsertion on the left and the number of served orders per reinsert step on the right.} \label{fig:reinsert_delay_plus_count}
\end{figure}

In Section \ref{sec:time_propagation} we distinguished between $\delta_{\text{delay,real}}$, the maximum delay allowed by the operator and $\delta_{\text{delay,heuristic}}$, the maximum delay used for running the proposed algorithm. Recall that this distinction is made to reduce the computational complexity and the time needed to solve the problem. By allowing reinsertions of orders, they have the chance to be served again, although with higher delay. How often each order can be reinserted is defined by Equation \ref{eq:reinsert}.\\
Here we set $\delta_{\text{delay,real}}=24$ minutes keeping $\delta_{\text{delay,heuristic}}=8$, which results in $\zeta=3$ maximum reinsertions per order. Figure \ref{fig:sensitvity_reinsert} shows a comparison with the baseline. We see an increase in service rate and a decrease in the total driven distance. All time KPIs increase. Increasing $\delta_{\text{delay,real}}$ allows for a higher delay per order, which leads to an overall higher average delay. The associated delay distribution is shown in Figure \ref{fig:reinsert_delay_plus_count}(a), where we observe a repetitive nature. The number of reinsertions is shown in Figure \ref{fig:reinsert_delay_plus_count}(b), showing a significant number of orders that are reinserted at least once.\\

\subsubsection{Number of orders $N$} 
\begin{figure}[H]
	\centering
	\includegraphics[width=1.0\textwidth]{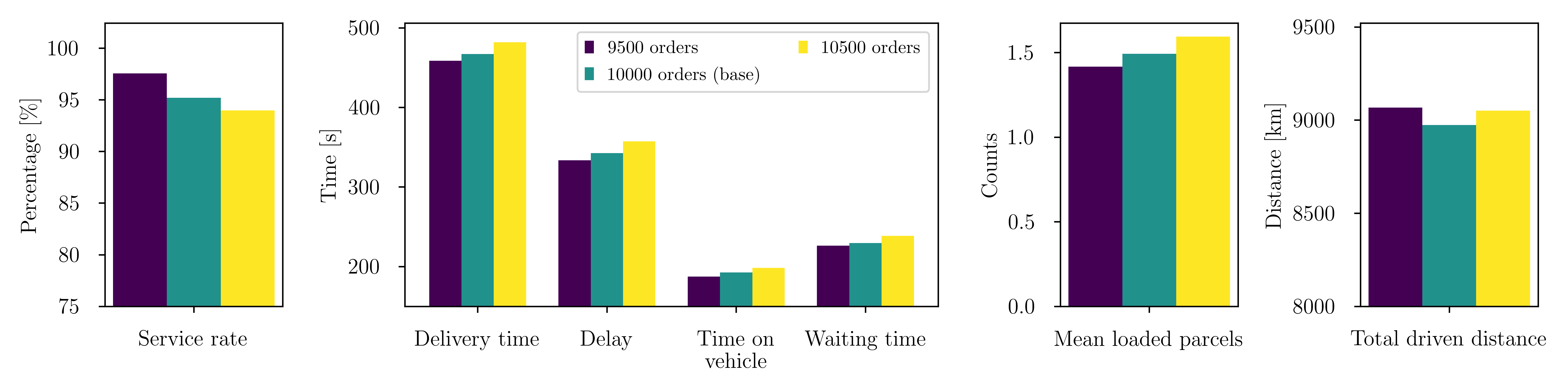}
	\caption{Service rate, time KPIs, mean loaded parcels, and total driven distance of the three different runs, featuring different demand patterns.}
	\label{fig:kpi_vs_demand}
\end{figure}

We created two alternative demand scenarios, featuring different numbers of customer orders $N$ (9,500 and 10,500) for the entire day. Both scenarios resemble the distribution of the 10,000 order case in time and space. Figure \ref{fig:kpi_vs_demand} depicts the corresponding results. The more orders are placed, the lower the service rate because available resources were kept constant. Nevertheless, the absolute number of delivered parcels increases (from 9,268 to 9,519, and then to 9,867). Although more parcels have been delivered, this does not necessarily increase the total driven distance. Compared with the 9,500 orders case, the driven distance decreases for both 10,000 and 10,500 orders. In general, if there are more orders to serve, it will be easier to find customer destinations close to each other, so the average distance between them decreases. However, those improvements do not hold for the time KPIs as all of them worsen with more placed and served orders. The number of mean loaded parcels increases in the same manner.

\subsubsection{Number of used vehicles:}
\begin{figure}[H]
	\centering
	\includegraphics[width=1.0\textwidth]{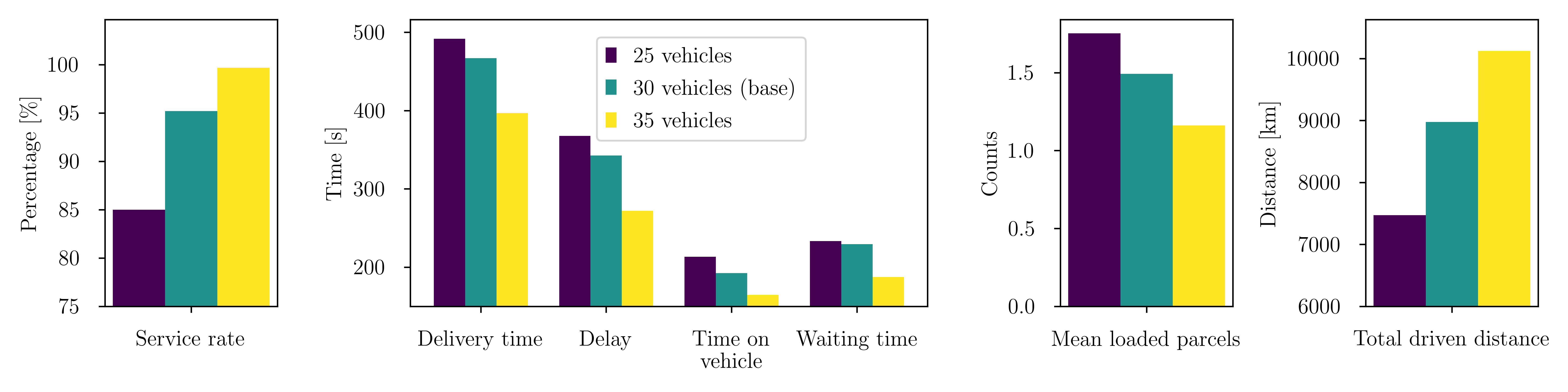}
	\caption{Service rate, time KPIs, mean loaded parcels, and total driven distance of the three different runs, featuring a different number of vehicles.}
	\label{fig:kpi_vs_vehicles}
\end{figure}

We varied the number of used vehicles to 25 and 35. Figure \ref{fig:kpi_vs_vehicles} shows the effect of the number of used vehicles on the obtained solutions. The service rate increases as more vehicles are used, and similarly, the total driven distance increases. Time KPIs decrease the more vehicles are used as well as the mean loaded parcels. The maximum travelled distance by one vehicle stays about the same, as those vehicles are utilized continuously throughout the whole day. For 25 vehicles, the maximal distance is 308\,km, for 30 vehicles is 311\,km, and for 35 vehicles is 309\,km. On the other hand, the minimal distance of one vehicle varies strongly. For 25 vehicles, the minimal distance is 290\,km, for 30 vehicles is 267\,km, and for 35 vehicles is just 228\,km. Thus, we conclude that with fewer vehicles, the vehicles are used closer to their full potential on average. When considering 35 vehicles some vehicles are used sparsely, especially at the start of the operation, where the overall workload is still low compared to later in the day.

\subsubsection{Cost function weight $\beta$:}
\begin{figure}[H]
	\centering
	\includegraphics[width=1.0\textwidth]{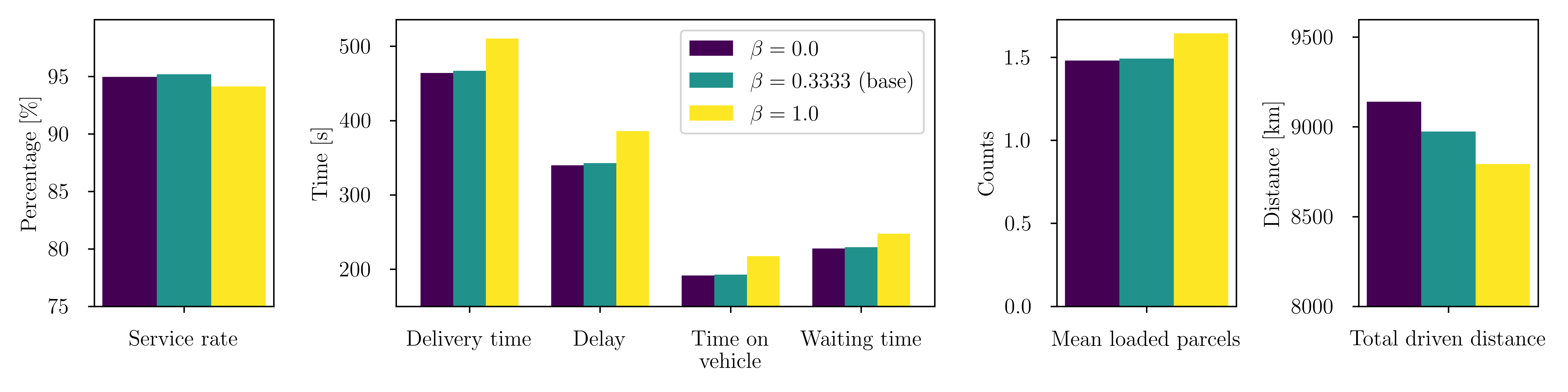}
	\caption{Service rate, time KPIs, mean loaded parcels and total driven distance of three runs, featuring different cost weights $\beta$.}
	\label{fig:kpi_vs_cost}
\end{figure}

Equations \ref{eq:cost_full} and \ref{eq:gamma} (global and one state) describe the cost function the method optimizes on. They are shaped by a parameter $\beta$, weighting service level and operational cost. $\beta=0$ leads to neglecting operational costs and the method fully trying to minimize delay, whereas the opposed happens for $\beta=1$. We run simulation for $\beta \in \{0, \frac{1}{3}, 1 \}$. Figure \ref{fig:kpi_vs_cost} displays the obtained results. For $\beta=0$ time KPIs are the lowest, as expected. For $\beta=1$, the total driven distance is minimized, as expected. The service rate varies slightly between the different runs. The highest service rate is achieved for $\beta=1/3$, which suggests that fully optimizing to minimize driven distance or delay leads the system into unfavourable states to serve future demand if the goal is to serve as many orders as possible.\\

\section{Conclusion} \label{sec:conclusion}
In this paper, we addressed the problem where a fleet of delivery vehicles serves a SDD operation, permitting to pick up goods at multiple depots and allowing vehicles to perform pre-empty depot returns, if beneficial. This enables a reduced average distance to customers' homes and more agile planning. We propose a method that works iteratively.
In each step, orders are assigned to potential pick-up locations, followed by checking how they could be combined into potential trips. As many trips as possible are generated for each vehicle, limited by predefined constraints on the total delivery times. Vehicles are assigned to the potential trips via solving an integer linear program.\\
The proposed method can handle large problem sizes. Extensive computational experiments simulating one day of service have been carried out. Looking at one scenario in detail, in which 10,000 orders are placed and 30 vehicles are available to serve those, a service rate of 95.19\% was achieved, which represents an improvement of 20\% over a greedy approach. The average delay accounts for 5\,min\,43\,s and 8,973.8\,km needed to be driven. 
Further, simulations showed the value of using and considering multiple depots and the value of performing pre-empty depot returns. A sensitivity study analyzed the varying influence of individual parameters on the obtained solution.\\
Future research could extend the proposed method to look ahead, such that the risk that the system gets into unfavourable states is reduced. Further, the possibility to plan for heterogeneous fleets of vehicles could be added. 
Additionally, the proposed approach is designed for on-demand deliveries exclusively. How to integrate already known orders is another future research question, such as the incorporation of stochastic information.
How many depots should be operated within a given area and how many of them should be considered per order are two further interesting question for the future.

% Acknowledgments here
\section*{Acknowledgement(s)}
This research was supported by Ahold Delhaize. All content represents the opinion of the author(s), which is not necessarily shared or endorsed by their respective employers and/or sponsors.

%%%%% APPENDIX %%%%%%%%%%%%%%%%
\appendix
\newpage
\textbf{\huge APPENDIX}
\section{Notation}
Table \ref{tb:notation} explains all notation used throughout this work.

\begin{longtable}[t!]{l|p{10.90cm}}
		\textbf{Variable} & \textbf{Description} \\
		\hline \hline
		\textbf{Environment} \\
		$G = (\mathcal{N}, \mathcal{A})$ & Weighted directed graph\\
		$\mathcal{N}$ & Set of nodes, each representing a specific location\\
		$\mathcal{A}$ & Set of weighted arcs, each connecting two nodes/locations\\
		$H$ & Number of depots present on the full graph $G$ \\
		$\mathcal{D}$ & Set of $H$ depots \\
		$d$ & Location of depot d\\
		
		\textbf{Vehicle fleet} & \\ 
		$v$ & Single vehicle\\ 
		$l_{v,t}$ & Location of vehicle $v$ at time $t$\\
		$\mathcal{LO}_{v,t}$ & Loaded orders of vehicle $v$ at timer $t$\\
		$M$ & Number of used vehicles \\
		$\mathcal{V}$ & Fleet of $M$ vehicles\\ 
		$C$ & Maximum capacity of a vehicle\\
		
		\textbf{Demand} & \\ 
		$o$ & Single order $o = \left\lbrace t_o, g_o \right\rbrace$\\
		$t_o$ & Time order o gets known\\
		$g_o$ & Goal location / destination of order o\\
		$d_{best,o}$ & The closest depot to an order's destination $g_o$\\
		$p_o$ & Pickup location of order o, not defined through the order o itself\\
		$N$ & Number of placed orders during one day of service \\
		$\mathcal{O}$ & Set of all $N$ orders $\mathcal{O}$ \\
		$\mathcal{UO}_t$ & Set of all orders that are not yet known (unknown orders)\\
		$\mathcal{PO}_t$ & Set of all orders that are known and not yet loaded by a (placed orders)\\
		$\mathcal{LO}_t$ & Set of all orders that are loaded by a (loaded orders)\\
		$\mathcal{DO}_t$ & Set of all orders that are already delivered (delivered orders)\\
		$\mathcal{IO}_t$ & Set of all orders that got ignored (ignored orders)\\
		
		\textbf{Times} \\
		$t$ & Current time\\
		$t_k$ & Time at decision k\\
		$T_{end}$ & End of the working day\\
		$\delta_{T}$ &time span, in which no more orders are placed\\
		$\delta_{load}$ & Loading time of one order at a depot\\
		$\delta_{service}$ & Service time for one order at its destination\\
		$\tau_{x_1,x_2}$ & Travel time between the two locations $x_1$ and $x_2$ \\
		$\delta_{delay}$ & Maximum delay until an order has to be delivered\\
		$t_{pick,o}$ & Pick-up time of order o\\
		$t_{drop,o}$ & Drop off time of order o\\
		$t_{drop,o,max}$ & Maximum drop off time of order o\\
		$t_{ideal,o}$ & Ideal (earliest) delivery time of order o\\
		
		\textbf{Candidates}\\
		$c$ & Candidate c, associated with order o, $c=(o_c,p_c)$ described through the order o itself and a potential pick-up location \\
		$\mathcal{C}$ & Set of all candidates \\
		$x$ & Number of depots considered per order as potential pick-up locations \\
		
		\textbf{Objective} \\
		$J$ & Cost function\\
		$\alpha$ & Penalty for not delivering an order\\
		$\beta$ & Weight in the cost function\\
		$\tau_{v}$ & Total travel time to serve all orders assigned to vehicle $v$\\
		
		\textbf{Miscellaneous} \\
		$P_v$ & Full day plan of vehicle $v$ \\
		$\Omega$ & Plans of all vehicles, which provides a solution to the problem, \quad \quad \quad \quad $\Omega = \cup_{v \in \mathcal{V}} P_v$ \\
		$\eta$ & Maximum number of orders in one trip \\
		$\theta_o$ & Delay of order o\\
	    $I^C_o$ & Set of candidates associated with order o\\ 
	    
	    \textbf{Splitting the problem} \\
		K & Total number of routing decisions during the course of one day\\
		k & Enumerates taken decisions\\
		$S_t$ & State at time $t$\\
		T & Trip, an ordered sequence of locations to visit\\

		$\mathcal{T}_k$ & Set of trips up to size k\\
		
		\textbf{Reinsertion of orders} \\
		$\delta_{\text{delay,real}}$ & Maximum delay set by the operator\\
		$\delta_{\text{delay,heuristic}}$ & Maximum delay the method uses in execution\\
		$\zeta$ & Limit of reinsertion, until an order is considered as ignored\\
		
	    \textbf{Assignment} \\
		$\epsilon_{\mathcal{T}\mathcal{V}}$ & Set of all feasible trip vehicle combinations\\
		$\epsilon_{\mathcal{T},v}$ & Corresponding binary variable \\
		$\mathcal{I}^T_{v}$ & Set of trips that can be serviced by a vehicle $v \in \mathcal{V}$\\
		$\mathcal{I}^T_c$ & Set of trips that contain candidate $c$\\
		$\mathcal{I}^{\mathcal{V}}_{T}$ & Set of vehicles that can service trip $T$\\
		$\mathcal{I}^{\mathcal{C}}_{o}$ & Set of candidates that belong to order $o$\\
		$\chi_o$ & Binary variable, taking the value of one if the corresponding order is ignored \\
		$\mathcal{X}$ & Set of all variables $\mathcal{X} = \{\epsilon_{\mathcal{T},v} , \chi_o ; \forall \epsilon_{\mathcal{T}\mathcal{V}} \text{ and } \forall c \in \mathcal{C} \}$ \\
		$\gamma_{T,v}$ & Costs of a vehicle's route \\
		$\gamma_{loaded,v}$ & Costs for the considered vehicle to serve its already loaded parcels are subtracted\\
		\\
	\caption{Notation and a short description of all used variables.}
	\label{tb:notation}
\end{longtable}

\section{Depot Distribution} \label{ap:k-center}
To locate the depots within the graph we used a k-center algorithm, which minimizes the maximum distance of all nodes to their closest depot. We used a greedy implementation of k-centers. It starts with a random node as the first depot, then iteratively, the node furthest from all depots is chosen to open a new depot. This is repeated until k depots are placed. We applied the algorithm 20 times using different starting points to account for the random starting node and choose the best set of depots.

\newpage
\section{Initialization Assignment Algorithm} \label{ap:algo}
\begin{algorithm}[h]
	\caption{Initialization Assignment}
	\label{alg:greedy_initalize}
	\SetKwInOut{Input}{input}
	\SetKwInOut{Output}{output}
	\Input{Set of all feasible trip vehicle combinations $\epsilon_{\mathcal{T}\mathcal{V}}$}
	\Output{Initialization assignment of trips to vehicles $\Omega_{\text{init}}$}
	\SetKwBlock{Beginn}{beginn}{ende}
	\Begin{
		$\mathcal{O}_{ok} = \emptyset \, \text{\scriptsize (Set of assigned orders)}$\; $\mathcal{V}_{ok} = \emptyset \, \text{\scriptsize (Set of assigned vehicles)}$\;
		\For{$ k = \eta$; $k \geq 0$; $k--$ $\text{\scriptsize \, starting at max trip size}$}{
			$L_k :=$ sort $\mathcal{T}_t$ in decreasing size\, $\text{\scriptsize (independent of vehicle)}$\;
			\If{size is equal}{sort by increasing cost $\gamma_{T,v} \, \text{\scriptsize (dependent on vehicle)}$ \;}
		}%\EndFor
		\ForEach{$T,v \in  L_k$}{
			\If{$(o(c) \notin \mathcal{O}_{ok} \, \, \forall c \in T) \textbf{ and } (r \notin \mathcal{V}_{ok}) \, \text{\scriptsize (none of the orders of the trip nor the vehicle are assigned yet)}$} {
				%\State
				$\mathcal{V}_{ok} \leftarrow v \, \text{\scriptsize (assign vehicle)}$\;
				$\mathcal{O}_{ok} \leftarrow o(c) \, \, \forall c \in T \, \text{\scriptsize (assign all orders of trip)}$\;
				$\Omega_{\text{init}} \leftarrow  T,v$\;
			}%\EndIf
		}%\EndWhile
		\Return $\Omega_{\text{init}}$
	}
\end{algorithm}
thereby, $o(c)$ gives back the order related to the candidate.

\section{Results} \label{ap:results_tabels}
The precise numbers of all performance indices of all executed runs are shown in Table \ref{tb:results_numbers}.\\
\begin{landscape}
\begin{table}[b]
\begin{tabular}{l|l|l|l|l|l|l|l}
	 & \textbf{\vtop{\hbox{\strut Service rate}\hbox{\strut [\%]}}} & \textbf{\vtop{\hbox{\strut Delivery}\hbox{\strut time [s]}}} & \textbf{\vtop{\hbox{\strut Delay}\hbox{\strut [s]}}} & \textbf{\vtop{\hbox{\strut Time on}\hbox{\strut vehicle [s]}}} & \textbf{\vtop{\hbox{\strut Waiting time}\hbox{\strut [s]}}} & \textbf{\vtop{\hbox{\strut Mean loaded}\hbox{\strut parcels}}} & \textbf{\vtop{\hbox{\strut Total distance}\hbox{\strut [km]}}} \\
	\hline \hline
	Base scenario & 95.19 & 467.07 & 342.61 & 192.53 & 229.54 & 1.49 & 8973.8\\
	
	\textbf{Comparison} & & & & & & \\
	\hline \hline
	Greedy & 74.18 & 533.93 & 406.30 & 184.81 & 304.12 & 1.13 & 10693.17\\
	1 depot per order & 92.68 & 456.74 & 332.82 & 177.23 & 234.51 & 1.36 & 9094.94\\
    No pre-empty & 94.81 & 470.89 & 346.43 & 191.73 & 234.16 & 1.48 & 8975.48\\	
	
	\textbf{Considered depots} & & & & & &   \\
	\hline \hline 
	5 depots per order & 94.95 & 472.08 & 347.54 & 198.28 & 228.80 & 1.53 & 9077.29\\
	7 depots per order & 95.25 & 467.12 & 342.66 & 195.60 & 226.52 & 1.52 & 9051.34\\
	
	\textbf{Total depots} & & & & & & \\
	\hline \hline 
	1 depot & 75.22 & 569.656 & 332.055 & 337.055 & 187.602 & 1.95 & 9555.17\\
	15 depots & 92.59 & 495.94 & 357.07 & 214.50 & 236.44 & 1.59 & 9714.92\\
    25 depots & 95.67 & 459.92 & 341.06 & 184.55 & 230.37 & 1.45 & 8844.39\\
	
    \textbf{Allowing reinserts} & & & & & & \\
	\hline \hline
    3 reinserts & 99.0 & 542.99 & 418.26 & 202.44 & 295.55 & 1.63 & 8803.1\\
    
	\textbf{Demand patterns} & & & & & & \\
	\hline \hline
	9500 orders & 97.56 & 458.64 & 333.60 & 187.37 & 226.27 & 1.42 & 9067.55\\
	10500 orders & 93.97 & 481.85 & 357.19 & 198.39 & 238.46 & 1.59 & 9050.93\\
    
    \textbf{Number of used vehicles} & & & & & & \\
	\hline \hline 
	25 vehicles & 84.98 & 491.69 & 367.70 & 213.30 & 233.39 & 1.75 & 7469.91\\
	35 vehicles & 99.67 & 396.95 & 272.14 & 164.58 & 187.37 & 1.16 & 10121.33\\

    \textbf{Cost weight} & & & & & & \\
	\hline \hline
    $\beta=0$ & 94.96 & 464.15 & 339.74 & 191.38 & 227.76 & 1.48 & 9140.61\\
    $\beta=1$ & 94.13 & 510.35 & 385.76 & 217.56 & 247.79 & 1.65 & 8793.51\\
\end{tabular} 
\caption{Precise results of all performance indices of all executed runs}
\label{tb:results_numbers}
\end{table}
\clearpage
\end{landscape}

\section{Base Scenario} \label{ap:base}
All parameter settings of the base scenario, analyzed in detail in Section \ref{sec:result_base}, are shown in Table \ref{tb:para_base}.\\
\quad
\begin{table}[h!]
	\begin{tabular}{l|l}
		\textbf{Parameter} & \textbf{Set value} \\
		\hline \hline
		$\delta_{\text{delay,real}}$, maximum Delay & 8 [min] \\
		$N$, number of vehicles & 30\\
		$C$, capacity per vehicle & 6 \\
		$H$, total number of depots & 20 \\
		$\beta$, weight in cost function & 0.33333 \\
		$\eta$, maximum size of a trip  & 10 \\
		$\zeta$, maximum number of reinsertion of an order & 1\\
		$\delta_{service}$, service time per order & 30 [s] \\
		$\delta_{load}$, loading time per order & 15 [s] \\
		$\Delta t$, time spans to run algorithm  & 100 [s] \\
		$\delta_{T}$, time in which no orders can be placed before the end of the day & 10 [min] \\
		$x$, number of depots considered per order & 3\\	
		vehicle speed & 10 [m/s]\\
		Trip generation timeout per vehicle $\rho_{max}$ & 50 [s]\\ 
	\end{tabular} 
	\caption{Parameter settings for the base scenario}
	\label{tb:para_base}
\end{table}

%%%%%%%%%%%%%%%%%%%%%% END APPENDIX %%%%%%%%%%%%%%%%%%%%%%%%%%%%%%

\bibliographystyle{apacite}
\bibliography{MyLibrary.bib}

\end{document}